\documentclass[prl,nobalancelastpage,twocolumn,superscriptaddress,showpacs,nofootinbib]{revtex4-1}

\usepackage{xcolor,amsthm,amsmath,amsfonts,graphicx,bm,amssymb,subfigure}
\usepackage{epstopdf}
\usepackage{ulem}
\usepackage{soul}

\newcommand{\MyTitle}{Magnetic crystals and helical liquids in alkaline-earth fermionic gases}

\date{\today}
\begin{document}

\title{\MyTitle}

\author{Simone Barbarino}
\email{simone.barbarino@sns.it}
\affiliation{NEST, Scuola Normale Superiore \& Istituto Nanoscienze-CNR, I-56126 Pisa, Italy}

\author{Luca Taddia}
\email{luca.taddia2@gmail.com}
\affiliation{Scuola Normale Superiore, 56126 Pisa, Italy}
\affiliation{CNR - Istituto Nazionale di Ottica, UOS di Firenze LENS, 50019 Sesto Fiorentino, Italy}

\author{Davide Rossini}
\email{davide.rossini@sns.it}
\affiliation{NEST, Scuola Normale Superiore \& Istituto Nanoscienze-CNR, I-56126 Pisa, Italy}

\author{Leonardo Mazza}
\email{leonardo.mazza@sns.it}
\affiliation{NEST, Scuola Normale Superiore \& Istituto Nanoscienze-CNR, I-56126 Pisa, Italy}

\author{Rosario Fazio}
\email{rosario.fazio@sns.it}
\affiliation{NEST, Scuola Normale Superiore \& Istituto Nanoscienze-CNR, I-56126 Pisa, Italy}

\begin{abstract}

The joint action of a synthetic gauge potential
and of atomic contact repulsion in a one-dimensional alkaline-earth(-like) fermionic gas 
with nuclear spin $I$ leads to the existence of a hierarchy of fractional
insulating and conducting 
states with intriguing properties. 
We unveil the existence and the features of those phases
by means of both analytical bosonization techniques and numerical methods 
based on the density-matrix renormalization group algorithm. 
In particular, we show that the gapless phases can support helical modes,
whereas the gapped states, which appear
under certain conditions,
are characterised both by density and magnetic order.
Several distinct features emerge solely for spin $I$ larger than $1/2$, thus making their study with cold-atoms unique. 
We will finally argue that these states are related to the properties of an unconventional fractional quantum Hall effect in the 
thin-torus limit. 
The properties of this hierarchy of states can be experimentally studied in state-of-the-art cold-atom laboratories.

\end{abstract}

\maketitle


The simultaneous presence of particle-particle interactions and of  (non-)Abelian gauge potentials, such as 
magnetic fields or spin-orbit coupling (SOC), is responsible for several spectacular phenomena, 
the fractional quantum Hall effect (QHE) being only the most known example~\cite{Tsui_1982}. 
Since several years this interplay is under close scrutiny both because of its perspective role in the realisation of 
robust quantum information protocols and because of the general interest in topological states of matter~\cite{Nayak_2008}. 
In the presence of SOC~\cite{Galitski_2013,ShizhongZhang_2014}, interactions can further drive the system into
fractional quantum spin Hall states~\cite{Kane_2005, Bernevig_2006_b} where edge-currents are spin polarised, or 
other exotic phases, characterised by unusual spin textures~\cite{Li_2013}, fractional conducting modes~\cite{Oreg_2014} or 
parafermions~\cite{Klinovaja_2014_b}.

Up to now, the attention has almost entirely focused on spin-1/2 (electronic) liquids, as most appropriate for the description 
of condensed matter systems. However, the recent progresses in the manipulation and control of cold atomic 
gases~\cite{Bloch_2008, Lewenstein_2007}  have brought to high relevance the study of systems of 
interacting fermions with a large (and tunable) spin.  
The investigation of alkaline-earth(-like) atoms such as Ytterbium~\cite{Sugawa_2011, 
Tale_2012, Pagano_2014, Cappellini_2014, Scazza_2014} or Strontium~\cite{Martin_2013, Zhang_2014}, which are characterised by a 
nuclear spin $I$ larger than $1/2$, is opening the path to the exploration of  phenomena, e.g. the properties of SU($\mathcal N$) 
models~\cite{Gorshkov2010, Cazalilla_2014}, which are not accessible with solid-state systems.  
The phase diagram of related multi-component Heisenberg or Hubbard-like models has been investigated in several 
works~\cite{Hermele_M_2009, Manmana_2011, Hermele_2011, Hazzard_2012, Messio_2012, Song_2013}; yet, very 
little is known concerning the effect of a gauge potential on an interacting system of particles with a large spin (see, however, Ref.~\cite{Chen_G}). 

Synthetic gauge potentials in cold atomic systems can be induced  via properly tailored laser pulses~\cite{Dalibard_2011}. 
The implementation of these schemes has already led to the realisation of light-induced magnetic fields~\cite{Lin_2009}, to 
Rashba SOC~\cite{Lin_2011}, as well as  to lattice models with non-zero Chern numbers~\cite{Tarruel_2012, 
Aidelsburger_2013, Miyake_2013, Jotzu_2014, Duca_2014, Aidelsburger_2014} and ladders with synthetic gauge 
potentials~\cite{Atala_2014}.  Similar approaches are suitable for application also to multi-component gases~\cite{Boada_2012, 
Celi_2014}, as shown in two recent spectacular experiments~\cite{mancini_2015, stuhl_2015}.
This fecund experimental activity, together with the rich scenario already explored for spin-$1/2$ systems, motivates the investigation 
of interacting systems with large spin coupled to gauge potentials. 

Do large $I$ alkaline-earth(-like) atoms lead to mere extensions of what is already known for electronic liquids? The answer is no. 
We will provide examples confirming that these setups allow for the exploration of novel regimes that can naturally be achieved 
only for $I > 1/2$ or through the versatility of this new playground. This puts cold-atom experiments in an excellent position for 
the investigation of novel intriguing many-body effects unattainable in conventional condensed matter setups. 

In this article we consider a one-dimensional fermionic gas with nuclear spin $I \geq 1/2$ and investigate the joint effect of 
interactions and of a synthetic gauge potential, which is 
equivalent to a Rashba SOC 
and an external magnetic field.
Provided that the states with highest and lowest spin are directly coupled with a multi-photon transition~\cite{Boada_2012, Celi_2014},
a full hierarchy of magnetic crystalline states appears at fractional fillings:
\begin{equation}
\nu \equiv \frac{ \pi n}{k_{\rm SO} \, (2I+1)} = \frac pq; \qquad p,q \in \mathbb N^+
\text{ and co-prime}
 \label{eq:condition}
\end{equation}
where $n$ is the atomic density and $k_{\rm SO}$ is the typical momentum of the SOC (to be defined in the following).
Combining analytical and numerical methods, we show that
these insulating phases exhibit charge and spin ordering and in some cases are connected to an unconventional fractional 
quantum Hall states in the thin-torus limit~\cite{Bergholtz2008}, while intrinsically resembling an edge of a spin Hall system.
The stabilisation of these gapped phases with $q>1$ requires some form of atom-atom interaction. Whereas some of 
the phases can be realised in the presence of a simple contact repulsion,  it is in general true the higher the $q$, 
the longer the range of the necessary interaction. 
These phases change dramatically once the mentioned multi-photon coupling is 
switched off: similarly to the spin-1/2 case~\cite{Oreg_2014}, fractional helical liquids appear.  

Our analysis, which includes also  the effect of a trapping potential, confirms that the findings of this work can be observed with 
state-of-the-art experimental techniques. 
Following the reasoning put forward in 
Ref.~\cite{Boada_2012, Celi_2014}, we can conclude that this setup might serve, especially in the limit of large $I$, as a quantum simulator of two-dimensional  
exotic interacting phases of matter in the strongly anisotropic limit.

\begin{figure}[t]
\begin{center}
	\includegraphics[width=\columnwidth]{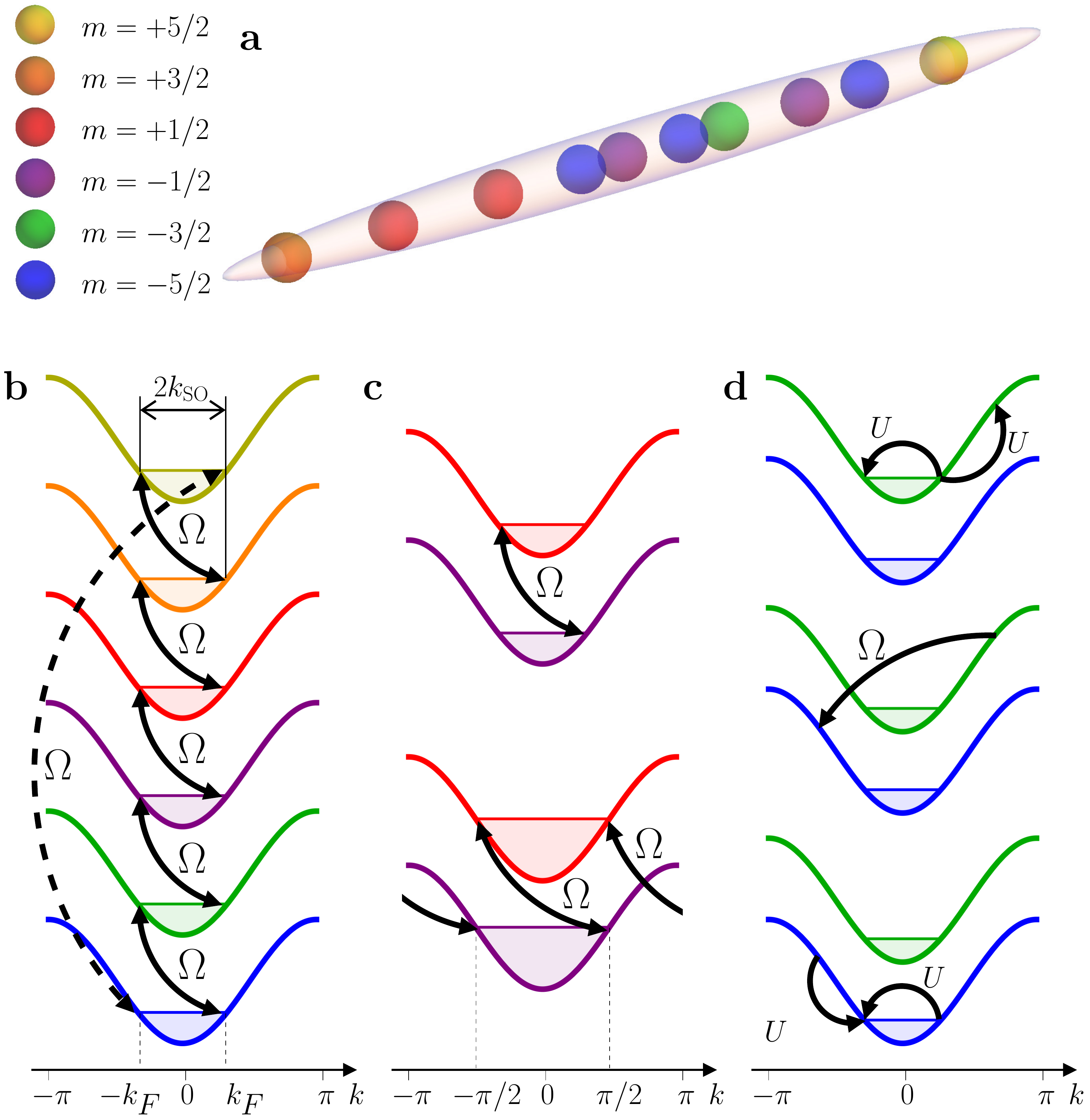}
\end{center}

 \caption{\textit{One-dimensional gas of $^{\rm 173}$Yb atoms.} (\textbf{a}) Sketch of a one-dimensional gas of $^{\rm 173}$Yb atoms with $I = 5/2$. 
 	(\textbf{b}) Energy bands of Hamiltonian~$\hat{\mathcal H}_0$ for $\hat {\mathcal H}_{\rm int}=0$ and $I=5/2$; an energy shift is 
 	inserted for representation clarity. When $\Omega$ is turned on,
	fermions with momentum difference $\Delta k =2 k_{\rm SO}$ 
 	and spin difference $\Delta m = \pm 1$ (solid arrows) or
 	$\Delta m = \pm 2I+1$ (dashed arrows) get coupled through 
 	$\hat {\mathcal H}_1$ and $\hat {\mathcal H}_2$, respectively:
 	if the  condition $k_{\rm SO} = k_F$ is met, the system develops a full gap, corresponding to $\nu = 1$.
 	(\textbf{c}) For $I=1/2$ the  condition $k_{\rm SO} = k_F$ is not enough
 	because $\hat {\mathcal H}_2=0$ (upper panel):
 	when $k_{\rm SO}=\pi/2$ the identification of momenta modulo $2 \pi $ allows for the creation of a gap (lower panel).
 	(\textbf{d}) When $\hat {\mathcal H}_{\rm int} \neq 0$, the system can develop a gap for lower fillings $\nu = 1/q$ via higher-order 
	scattering terms.  
	As an example, the picture highlights three intermediate processes: their sequence (top to bottom) originates a third-order process which couples two Fermi surfaces  with $\Delta m = \pm 1$ 
	for $q = 3$ and $k_F = k_{\rm SO}/3$.
 }
 \label{fig:Sketch1}
\end{figure}

\paragraph{ \textit{Model.}} 
We consider a one-dimensional model that describes an optical lattice loaded with a gas of fermionic alkaline-earth(-like) 
atoms whose ground state is characterised by a spin-$I$ nuclear manifold; for a sketch, see Fig.~\ref{fig:Sketch1}(a). 
Disregarding for the moment the harmonic confinement, the Hamiltonian reads~\cite{Gorshkov2010}
\begin{equation}
	\hat{\mathcal{H}}_0=-t\sum_{j} \sum_{m=-I}^I \left(
	\hat c_{j,m}^\dagger \hat c_{j+1,m}+\text{H.c.}\right)+\hat {\mathcal H}_{\rm int} \, ;
\label{H}
\end{equation}
where $\hat c_{j,m}^{(\dagger)}$ are fermionic operators annihilating (creating) an atom at site $j$ with nuclear spin $m$ 
and $t$ is the hopping amplitude. $\hat{\mathcal H}_{\rm int}$ describes an SU($2I+1$) invariant interaction which is
usually of contact kind: $ U\sum_j \sum_{m < m'} \hat n_{j,m} \hat n_{j, m'}$. As new proposals make the engineering of 
longer-range interactions in cold gases more practicable~\cite{Rydberg_2012}, we also discuss as an example
the effect of a nearest-neighbour potential:
$ V\sum_{j,m,m'} \hat n_{j,m} \hat n_{j+1, m'}$.

A Raman coupling endowed with a running phase  connects states which differ for one nuclear magnetic quantum $\Delta m = \pm1$: 
\begin{equation}
	\hat{\mathcal{H}}_1=\sum_{j} \sum_{m=-I}^{I-1} \left( \Omega_m e^{-i 2 k_{\rm SO} j} \hat c_{j,m}^\dagger 
	\hat c_{j,m+1}+\text{H.c.}\right);
\label{H}
\end{equation}
here, $\Omega_m = \Omega g_m$, with $\Omega$ the Raman-coupling strength and $g_m = \sqrt{I(I+1)-m(m+1)}$; see Ref.~\cite{Celi_2014} for a derivation of~\eqref{H}.

The unitary transformation $\hat {\mathcal U}$ defined by $\hat {\mathcal U} \hat c_{j,m} \hat {\mathcal U}^\dagger 
= e^{i 2 k_{\rm SO} m j} \hat c_{j,m}$ maps Hamiltonian $\hat {\mathcal H}_0+\hat {\mathcal H}_1$  to a spin-$I$ fermionic model in 
the presence of Rashba SOC and of a magnetic field $\Omega$ with perpendicular quantization axis (see Supplementary Material).
The choice to denote the phase factor in Eq.~\eqref{H} with $k_{\rm SO}$ becomes then clear upon inspection of the kinetic term: 
$\hat {\mathcal U} \, 
\hat {\mathcal H}_0 \, 
\hat {\mathcal U}^\dagger 
=- 2t \sum_{k,m} 
\cos (k-2m k_{\rm SO})
\hat c_{k, m}^\dagger 
\hat c_{k, m} +
\hat {\mathcal H}_{\rm int} $.

Alternatively, $\hat {\mathcal H}_0+\hat {\mathcal H}_1$ 
can be interpreted as a spinless fermionic model with 
one (finite) synthetic dimension coupled to a synthetic Abelian 
gauge field~\cite{Celi_2014}.
Multi-photon transitions can implement an additional coupling between the extremal nuclear spins $m = \pm I$, described by  a contribution $\hat{\mathcal{H}}_2=\sum_j \left( \Omega' e^{-i2 k_{\rm SO} j} \hat c_{j,I}^\dagger \hat c_{j,-I}+\text{H.c.}\right)$ 
to the Hamiltonian~\cite{klinovaja_2012}.
In the following we consider $\Omega'=\Omega$ because, 
especially for small $I$,
multi-photon transitions $\Omega'$ can be tuned to the order of magnitude of the Raman-coupling $\Omega_m$.
The presence or absence of $\hat{\mathcal{H}}_2$ corresponds to different boundary  conditions 
along the synthetic dimension, a feature that is relevant only for $I > 1/2$ 
because, for $I=1/2$, Hamiltonian $\hat {\mathcal H}_2$ connects states that are already linked by $\hat {\mathcal H}_1$.
The possibility to tune (and eventually switch off) $\hat{\mathcal{H}}_2$ is unique to this cold-atom implementation.

\begin{figure*}[t]

\includegraphics[width=\columnwidth]{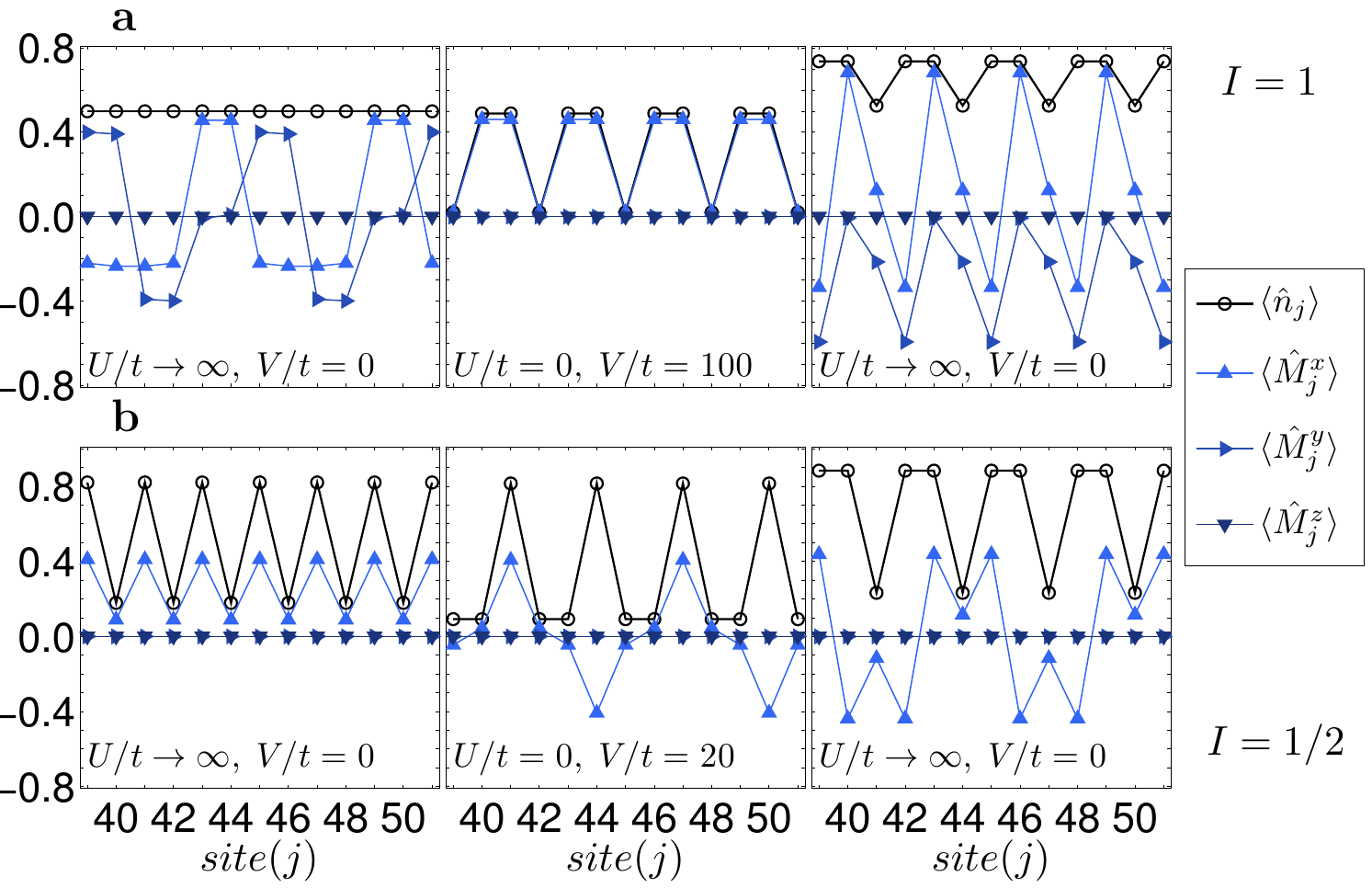}
\includegraphics[width=\columnwidth]{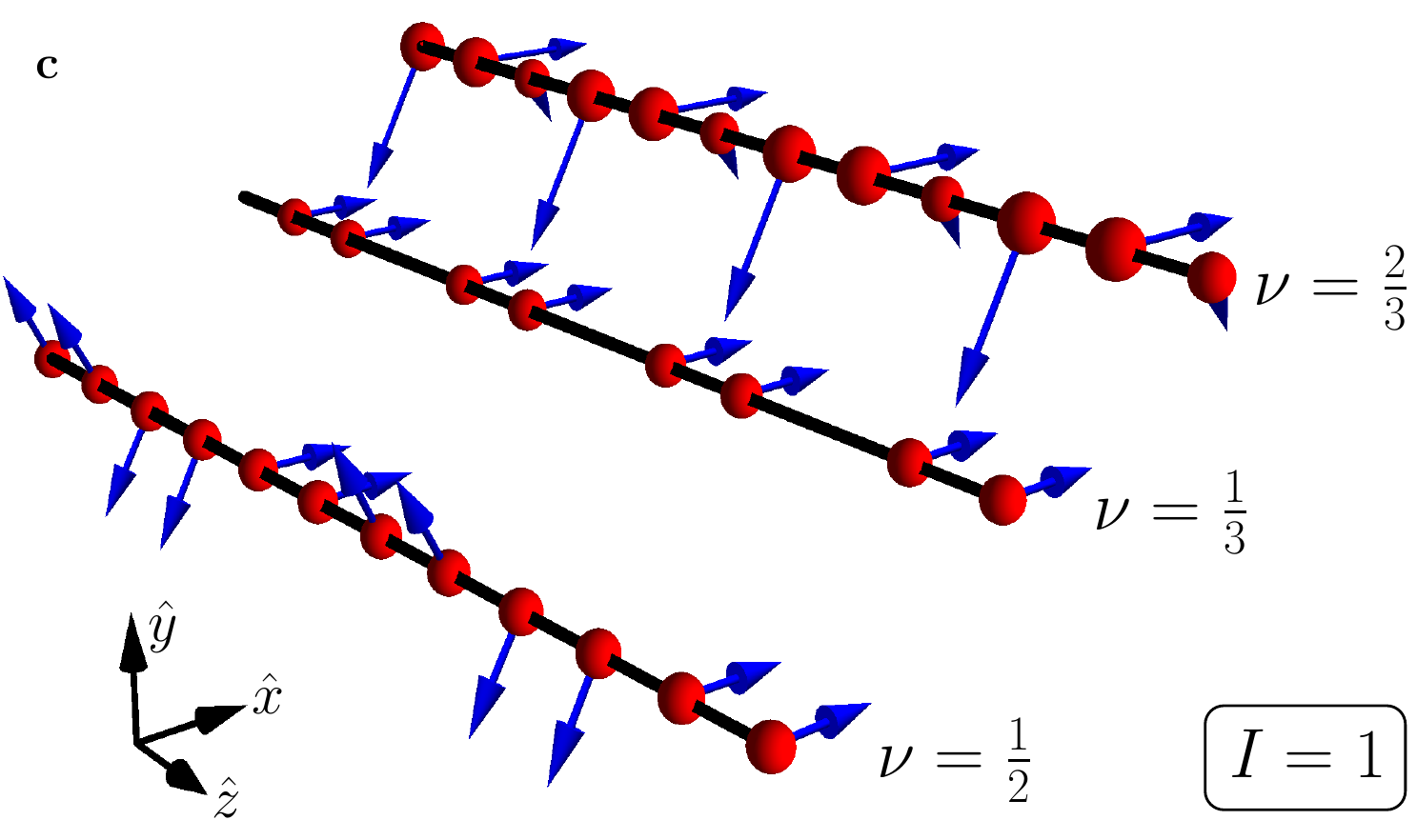}

\caption{
	\textit{Charge and magnetic order of the magnetic crystals.} Charge $\langle \hat n_i \rangle$ 
	and magnetic order $\langle \hat M^\alpha_i \rangle$ of the fractional phases at
	$\nu = \frac 12$, $\frac 13$  
	and $\frac 23$ (from left to right)
	for $I=1$, $k_{\rm SO} = \pi /3$ (\textbf{a}) and
	$I=1/2$, $k_{\rm SO} = \pi/2 $ (\textbf{b})
	as obtained from DMRG simulations 
	of a system of length $L=96$ with $\Omega/t = 1$.
	For the interaction parameters see the panels.
	Since the system is a crystal with small boundary effects, we only plot the central part of the system for a better readability.
	(\textbf{c}): Sketch of the density (red balls with radius related to $\langle \hat n_j\rangle$) and magnetic properties (blue arrows representing the vector $\langle \hat M_j^\alpha\rangle$) of the 
	insulators at $\nu = \frac 12$, $\frac 13$ and $\frac 23$ (from left to right) for $I=1$.
 }
\label{fig:Charge:Magnetic}
\end{figure*}

Finally, the presence of a trapping potential is taken into account through an additional term in the Hamiltonian of the form
$\hat{\mathcal{H}}_{\rm tr}=\sum_{j,m} w_j \, \hat c_{j,m}^\dagger \hat c_{j,m}$.
The model defined by the Hamiltonian $\hat {\mathcal H}_0+\hat {\mathcal H}_1 + \hat {\mathcal H}_2 + \hat{\mathcal{H}}_{\rm tr}$ will 
be studied in its various limits by means of bosonization techniques~\cite{Giamarchi2004} and density-matrix renormalization group (DMRG) simulations~\cite{Schollwoeck_2005, Schollwoeck_2011}, 
which are specially suited for interacting systems in one dimension.

\paragraph{ \textit{ Magnetic crystals.} }
As a  first step we show that for the fillings given in Eq.~\eqref{eq:condition} the system described by  Hamiltonian $\hat {\mathcal H}_0+
\hat {\mathcal H}_1+\hat {\mathcal H}_2$ and with $I \geq 1$
is insulating. Note that there is no inconsistency between the 
fermionic statistics and the study of integer $I$, as the considered states can be selected from a larger half-integer manifold~\cite{Gorshkov2010, Pagano_2014}.
The peculiar case $I=1/2$ is also considered.

Gapped phases appear  in the absence of interactions only for the fractions in Eq.~\eqref{eq:condition} with $q=1$. It is  
instructive to briefly discuss this case as it helps understanding the interacting one. By writing Hamiltonian $\hat {\mathcal H}_0+
\hat {\mathcal H}_1+\hat {\mathcal H}_2$ in momentum-space representation  and defining a Fermi energy and a Fermi momentum 
$k_F$ relative to the case with $\Omega=0$, it is immediate to see that whenever $k_{\rm SO} = k_F$,  a fermion can be spin-flipped 
and  scattered from one edge of the Fermi sea to the other one.  As any edge of the Fermi sea gets coupled, the system develops a 
full gap (see Fig.~\ref{fig:Sketch1}(b)). In this case, the density $n$ is $(2I+1)  k_F /\pi$, so that $\nu = 1$. Because with higher-order processes 
$\hat {\mathcal H}_1$ and $\hat {\mathcal H}_2$ can connect spin states with $|\Delta m| >1$,  a gap opens whenever $k_{\rm SO} =  k_F /p$, $p \in [1,2 \ldots 2I]$:
these phases correspond to $\nu = p$.

The case $I=1/2$ is peculiar since $\hat {\mathcal H}_2=0$ and  only two edge-modes gap out~\cite{Braunecker_2010}.
Exploiting the presence of a lattice, it is however possible to invoke the equivalence of momenta upon shifts of $2 \pi$ and 
hence to derive the additional condition $k_F + 2 k_{\rm SO} = -k_F+ 2\pi$. Together with $k_{\rm SO} = k_F$ (in this case $k_{\rm SO} =  \pi/2$), it ensures 
that the system enters a gapped phase at $\nu = 1$ (see Fig.~\ref{fig:Sketch1}(c)).
Contrary to the previous case, 
the resulting gapped phase is a lattice effect, without a proper continuum limit. 

The existence of gapped phases for $q>1$ and odd can be predicted using the bosonization technique~\cite{Giamarchi2004, Kane_2002, Teo_2014}, 
which is the natural analytical tool to take into account interactions in one dimension (see the Supplementary Material for a discussion 
of the case of even $q$, together with more details about the bosonization approach). By linearizing the non-interacting spectrum 
of $\hat {\mathcal H}_0$ around the Fermi energy, we can write $\hat c_m(x)= e^{i k_Fx} \hat \psi_{+,m}(x) + e^{-ik_Fx} \hat \psi_{-,m}(x)$, where
$\hat \psi_{+,m}$ ($\hat \psi_{-,m}$) is the right (left)-moving operator of the $m$-th nuclear spin state. Central to the theory 
is its expression in terms of the bosonic fields $\hat \phi_{r,m}(x)$, $r= \pm$, as $\hat \psi_{r,m}(x) \sim \exp{[-ir \hat \phi_{r,m}(x)]}, $
 with $[\hat \phi_{r,m}(x), \hat \phi_{r',m'}(x')] = - i \, r \, \pi \, \mathrm{sgn}(x'-x) \, \delta_{r,r'} \, \delta_{m, m'}$.
In the presence of density-density interaction terms, such as  $\hat {\mathcal H}_{\rm int}$,  
 the full Hamiltonian $\hat {\mathcal H}_0$ can be cast into the quadratic form
\begin{equation}
	\hat{\mathcal{H}_0} =\sum_{m,m'} \int \;dx\ 
	(\begin{matrix} \partial_x \hat{\varphi}_m(x) & \partial_x \hat{\theta}_m(x) \end{matrix} )
	\mathcal{M}_{m,m'}
	\left(\begin{matrix} \partial_x \hat{\varphi}_m(x) \\ \partial_x \hat{\theta}_m(x) \end{matrix} \right) \label{HamLL}
\end{equation}
with $\mathcal{M}_{m,m'}= v_F /(2\pi) \delta_{m,m'}+ \mathcal{U}_{m,m'}$; $v_F$ is the Fermi velocity, and $\hat \phi_{r,m}(x) = 
\hat{\varphi}_m(x) - r \hat{\theta}_m(x)$; $\mathcal{U}_{m,m'}$ describes the scattering processes induced by $\hat {\mathcal H}_{\rm int}$ only.

Insulating phases are determined by the joint action of  
$\hat{\mathcal H}_1+\hat{\mathcal H}_2$ and $\hat {\mathcal H}_{\rm int}$, representing the interplay of the gauge potential and interactions. 
Indeed, $\hat{\mathcal H}_1$ originates $2I$ processes of the form
\begin{equation}
	(\hat \psi^\dagger_{+,m} \hat \psi_{-,m})^n 
	\; \hat \psi^\dagger_{+,m} \hat \psi_{-,m+1} 
	\; (\hat \psi^\dagger_{+,m+1} \hat \psi_{-,m+1})^{n},
	\quad n \in \mathbb N^+;
\label{sinegordon}
\end{equation} 
with $m=-I,...,I-1$,
which are relevant only when momentum is conserved, namely $k_F=k_{SO}/q$ with $q=2n+1$. 
When $n=1$, for instance, term~\eqref{sinegordon} reproduces the low-energy physics of the processes displayed in Fig.~\ref{fig:Sketch1}(d).
$\hat{\mathcal H}_2$ generates one 
more relevant process, expressed in the form of Eq.~\eqref{sinegordon} by replacing $m$ with $I$  and $m+1$ with $-I$. 
Once Eq.~\eqref{sinegordon} is written in terms of the bosonic fields, $2I+1$ Sine-Gordon Hamiltonians with commuting arguments 
are obtained.  Following Ref.~\cite{Kane_2002, Teo_2014}, when these terms are relevant in a renormalization group sense,  
they lead to the formation of $2I+1$ mass gaps which make $\hat {\mathcal H}_0+\hat {\mathcal H}_1+\hat {\mathcal H}_2$ fully gapped.

The request for relevance in the renormalization group sense
points out an important fact: given a specific filling, 
a generic interaction does not necessarily stabilize a gapped phase.
Through the explicit study of the $I=1/2$ case, for
which precise mappings between the microscopic interacting model and 
the bosonization parameters are known, in the Supplementary Material
we argue that the higher the value of $q$, 
the longer the range of the interactions required to make the 
$2I+1$ Sine-Gordon terms relevant.

In order to fully characterise the properties of this hierarchy of gapped phases, we perform numerical simulations with 
DMRG~\cite{Schollwoeck_2005, Schollwoeck_2011} that in one dimension provide essentially exact results  and  allow us to explore the cases 
of even $q$ and $p > 1$, which are not easily accessible with bosonization (see the Supplementary Material for details on the numerical simulations).  
We particularly focus on  $I = 1/2$ and $I=1$ 
(for these cases $g_m =g$ and can be set to $1$)
and use the latter case to highlight the general new features emerging for $I> 1/2$.

The results in Fig.~\ref{fig:Charge:Magnetic} consider the  fillings $\nu = \frac 12$, $\frac 13$ and $\frac 23$ and display the density profile $\langle \hat n_j \rangle \equiv \sum_m 
\langle \hat n_{j,m} \rangle$ and the magnetization $\langle \hat M^\alpha_j \rangle \equiv \sum_{m,m'} \langle 
\hat c^\dagger_{j,m} [S^{\alpha}]_{m,m'} \hat c_{j,m'} \rangle$, where $S^{\alpha}$ is a $(2I+1 )$-dimensional 
representation of the SU($2$) spin operator ($\alpha =x$, $y$, $z$).  
Fig.~\ref{fig:Charge:Magnetic} clearly shows that the incompressible phases are characterised by both charge and magnetic order.

Let us begin with the charge order, considering for example the case $I=1/2$, $\nu = \frac 13$; here, a nearest-neighbour 
interaction stabilises a density wave with one-particle every three sites. Note that in the usual SU(2) Hubbard model ($\Omega = 0$, 
no gauge potential) a nearest-neighbour interaction only stabilises a density wave with one particle  every two sites. 
Conversely, for $U,V=0$ (no interactions) no gapped phases exist at this filling.
The simultaneous presence of interactions and of the (commensurate) SOC proves to be crucial in all the considered cases
both for the opening of the gap and  for crafting the properties of the resulting  insulator. 

Concerning magnetic order, in all the considered cases $\langle \hat M^z_j\rangle=0$, which implies that in every site the $\pm m$ 
nuclear-spin components have the same occupation numbers. Magnetization  lays in the $x$---$y$ plane and winds as a function of position in peculiar 
ways which depend on the filling (see the sketches in Fig.~\ref{fig:Charge:Magnetic} for $I=1$). For $I=1/2$, 
the Hamiltonian is real and thus $\langle \hat M^y_j\rangle=0$: the magnetization only develops along the $\hat x$ axis.

\begin{figure}[b]

\includegraphics[width=\columnwidth]{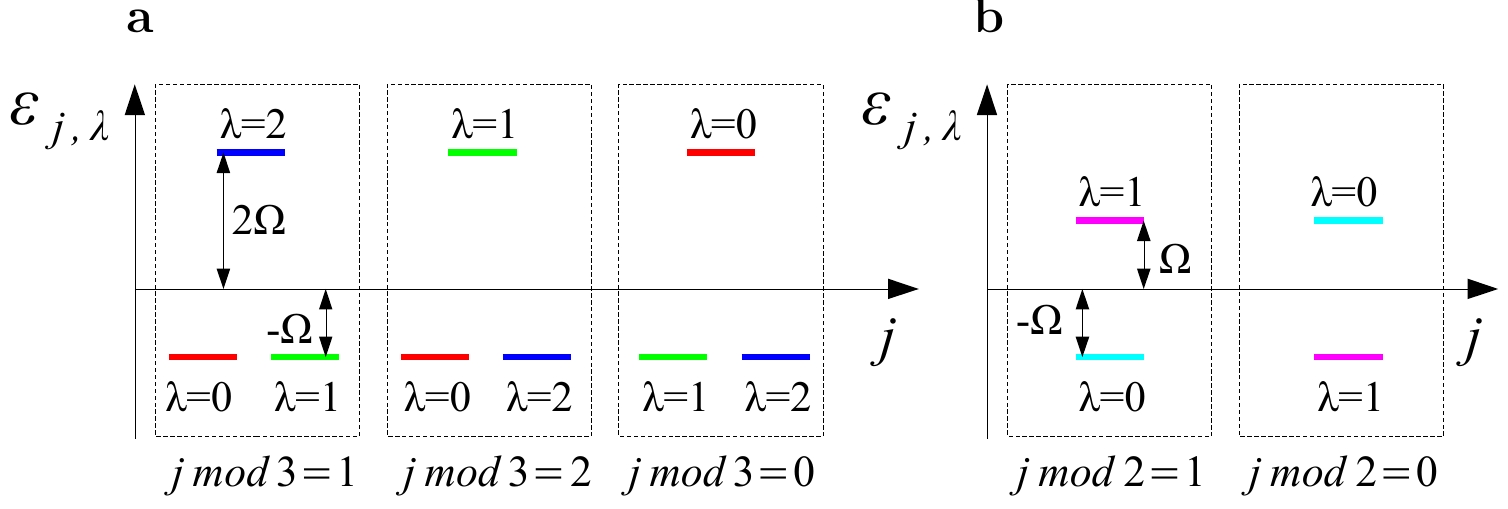}

\vspace{0.2cm}

\includegraphics[width=\columnwidth]{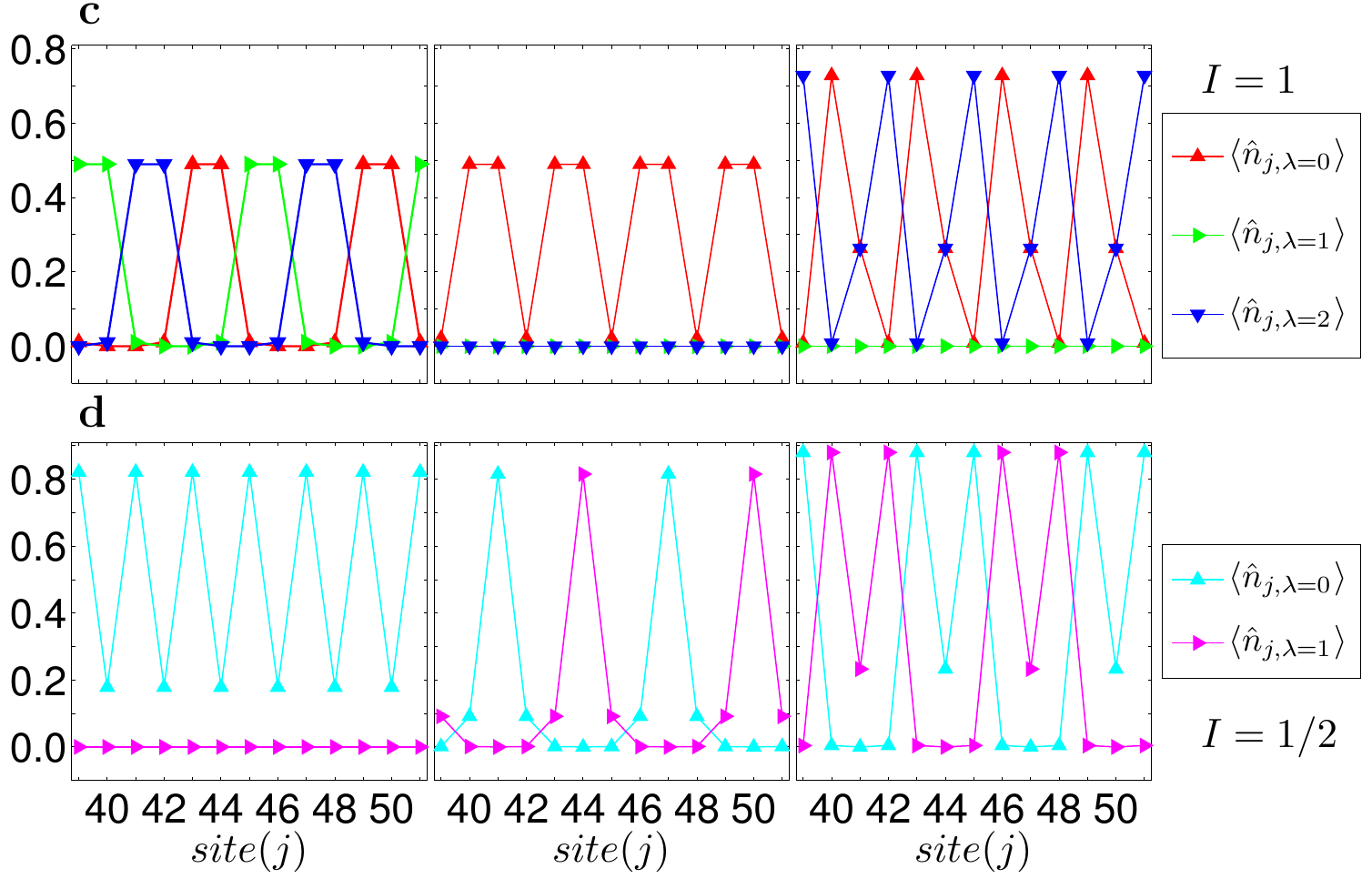}
 \caption{
 	\textit{Magnetic crystals.} Energy-spin structure $\varepsilon_{j,\lambda}$ 
 	for: (\textbf{a}) $I=1$ and $k_{\rm SO} = \pi/3$,
 	(\textbf{b}) $I = 1/2$ and $k_{\rm SO} = \pi/2$.
 	Density plots $\langle \hat n_{j,\lambda} \rangle \equiv \langle \hat d_{j,\lambda}^\dagger \hat d_{j,\lambda}\rangle$ 
 	at $\nu = \frac 12$, 
 	$\frac 13$ and $\frac 23$ (from left to right)
 	for $I = 1$ (\textbf{c}) and $I=1/2$ (\textbf{d}) obtained through DMRG simulations (see the caption of Fig.~\ref{fig:Charge:Magnetic} for the parameters).
 }
\label{fig:SU3:SU2:PBC}
\end{figure}

The density and magnetic ordering found in DMRG simulations becomes transparent by considering the limit $\Omega / t \gg 1$, $\Omega/U \gg 1$ where a 
simple model catches its salient features. 
Hamiltonian $\hat {\mathcal H}_1+\hat {\mathcal H}_2$  
is invariant under the $\mathbb Z_{2I+1}$ group related to the on-site transformation $m \mapsto m+1$ (here, $I+1 \equiv -I$), and is 
diagonalized by $2I+1$ local eigenmodes of the form: 
\begin{equation}
 	\hat d_{j,\lambda} = \frac{1}{\sqrt{2I+1}}\sum_m
	\omega^{\lambda m} \hat c_{j,m};
	\quad
	\omega = e^{i 2 \pi /(2I+1)};
\end{equation}
with $\lambda = 0 \ldots 2I$.  As $\hat {\mathcal H}_0$ is SU($2I+1$) invariant, it is unaffected by this basis change.
However, the on-site energy of each transformed spin state  depends both on the site $j$ and on $k_{\rm SO}$:
$\varepsilon_{j,\lambda} = 2 \Omega \cos \left[ \frac{2 \pi \lambda} {2I+1} +2 k_{\rm SO} j \right]$. 
When the atomic density $n$ and $k_{\rm SO}$ are commensurate, the periodic energy landscape $\varepsilon_{j,\lambda}$ induces charge-density waves, with atoms magnetised accordingly.
As shown in Fig.~\ref{fig:SU3:SU2:PBC}, this pattern ultimately determines the charge and spin order
even at $\Omega/t \sim 1$.
For $I=1$ and $k_{\rm SO}=\pi/3$, the system dimerizes (this is because $\varepsilon_{j,\lambda}$ has two degenerate lowest-energy states) and for 
$\nu = \frac12$ and $\frac 13$ dimers arrangements are quite apparent. For $I=1/2$, the modes $\hat d_{j, \lambda}$ are 
the eigenstates of $\hat M^x_j$, and this explains why magnetization develops only along $\hat x$.
This simple model provides a physical intuition of the fact that $\hat {\mathcal H}_1+\hat {\mathcal H}_2$ induces magnetic crystals
at the fillings in Eq.~\eqref{eq:condition}, thus complementing  the information provided by bosonization for odd $q$ and by DMRG
for the other considered cases.

\begin{figure}[t]
\includegraphics[width=\columnwidth]{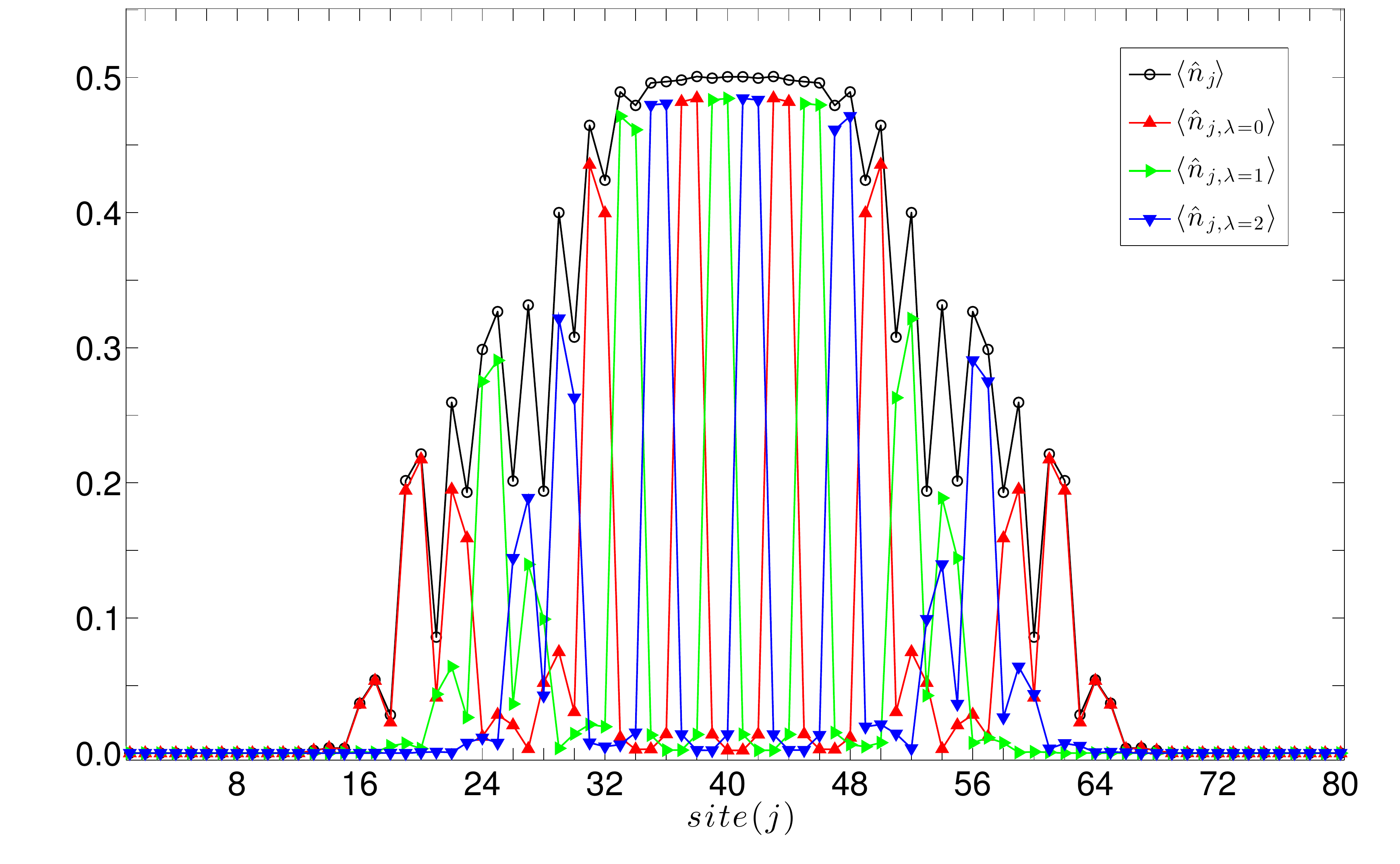}
\caption{\textit{Harmonic confinement effects.} Density profile of the system $I=1$ in the presence of a harmonic confinement with $\bar w = 10^{-3}$. The parameters of the simulation, done for $16$ fermions, are  $U/t =10 $, $\Omega/t=1$ and $k_{\rm SO} = \pi/3$. In the center of the system the typical charge and magnetic order of the insulator with filling $\nu = \frac 12$ appear.}
\label{sec:trap}
\end{figure}

The gapped phases in Fig.~\ref{fig:SU3:SU2:PBC} at $\nu = \frac12$ and $\frac 23$ are stabilized by the solely on-site repulsion, and thus potentially experimentally 
accessible in current experimental setting. It is therefore important to check whether the presence of the trapping potential will modify 
our findings. This is explored in Fig.~\ref{sec:trap} where we show  that the presence of a harmonic confinement,  introduced through 
$\hat {\mathcal H}_{\rm tr}$ with $w_j =  \bar w \, (j-L/2 -1/2)^2$ does not hinder the possibility of observing the peculiar properties  of the 
magnetic crystals discussed so far; in a Thomas-Fermi spirit, they form in definite regions of the trap. It is natural to envision that the 
peculiar density-spin patterns which characterise these gapped phases could be unambiguously revealed through spin-resolved single-site 
addressing~\cite{Bakr_2009, Sherson_2010}. Alternatively, Bragg scattering could get access to the structure factor of the gas, which 
contains information about the density order and, if spin-resolved, even about the magnetic one~\cite{Corcovilos_2010}.

\paragraph{ \textit{ Helical liquids.} }

According to the previous discussion, for $I \geq 1$ gapped phases 
at the fillings in Eq.~\eqref{eq:condition} arise in the interacting system only when 
$\hat {\mathcal H}_2 \neq 0$ and, for $I=1/2$, when $k_{\rm SO} = \pi /2$ (lattice effect).
More generally, explicit inspection has shown that even when $\hat {\mathcal H}_2=0$
the condition $k_{\rm SO} = \pi/(2I+1)$ for $I \geq 1$ can open 
a full gap through a high-order process which exploits the presence
of a lattice (i.e. of momenta identification upon shifts of $2 \pi$).
When these conditions are not met, and thus the full gap does not develop,
we will show that 
for the fillings in Eq.~\eqref{eq:condition} the system described by
$\hat {\mathcal H}_0+\hat {\mathcal H}_1$ is a helical liquid~\cite{Oreg_2014}. 
 
An intuition for this phenomenology 
is again provided by the non-interacting case $\hat{\mathcal H}_{\rm int}=0$ and $q = 1$. Here, the two Fermi edges at 
$m=\pm I$ and $k = \pm k_F$, respectively, remain unperturbed (see Fig.~\ref{fig:Sketch1}(b)) and represent two gapless helical 
fermionic modes, which are the lowest-energy excitations of the system. Bosonization techniques applied to the case  $\hat {\mathcal H}_{\rm int} \neq 0$
for $q \geq 1$ and odd clearly pinpoint the helical nature of the low-energy spectrum, although the gapless modes for $q>1$ are  linear combinations of the original modes. Additionally, their conductance is fractional~\cite{Oreg_2014}: these two properties define a fractional helical liquid.
Similarly to the magnetic crystals, bosonization reveals the existence of requirements on the range and intensity of the atom-atom repulsion.

Numerically we cannot fully access the helical nature of the first excitations; we rather diagnose the existence of a helical ground state through two observable quantities. 
First, as a consequence 
of the existence of two gapless modes, the low-energy  spectrum is described by  a  conformal field theory in 1+1 dimensions with 
central charge $c=1$. Second, the system is characterised by a current  pattern $\langle \hat {\mathcal J}_{j,m}^z  \rangle$,
where $\hat {\mathcal J}_{j,m}^z \equiv -i t \hat c^{\dagger}_{j,m} \hat c_{j+1,m} + \mathrm{H.c.}$,
which is different from zero and features a flow direction related to the sign of $m$.

Our numerical results for the case $I=1$, $k_{\rm SO} = \pi/3$ 
and $\nu = \frac 23$
are shown in Fig.~\ref{fig:Helical:OBC} (for other fillings and values of $I$ see the Supplementary Material).
For ground states of theories with a low-energy conformal limit, the Calabrese-Cardy formula~\cite{Calabrese_2004} predicts that
the entanglement entropy $S(\ell) =-\text{Tr}[\rho_\ell \log \rho_\ell]$, obtained through a bipartition  of the system into two blocks of length $\ell$ and $L-\ell$ ($\rho_\ell$ is the reduced density matrix of the block of size $\ell$), should be proportional to $c$: $S(\ell) = A + c/6 \log [2L/\pi \sin (\ell \pi / L)]$ (for open boundary conditions). Figure~\ref{fig:Helical:OBC} shows $S(\ell)$ 
as computed through DMRG simulations for $L=192$ and a fit with the previous formula  leaving $A$ and $c$ as free parameters.
The result clearly indicates $c=1$. 
The current profile $\langle \hat {\mathcal J}_{j,m}^z  \rangle$ displayed in Fig.~\ref{fig:Helical:OBC}
shows a finite saturated value for $j \sim L/2$ and $m = \pm 1$ (note that they flow in different directions);
for $m=0$ the absence of a current is imposed by symmetry reasons. The oscillations are a boundary and finite-size effect  
and decay to zero for $L\to \infty$ (see Supplementary Material). 
Note that in this case, even if $k_{\rm SO} = \pi / (2I+1)$, the system is gapless and thus amenable to the study of helical properties.

\begin{figure}[b]

\includegraphics[width=\columnwidth]{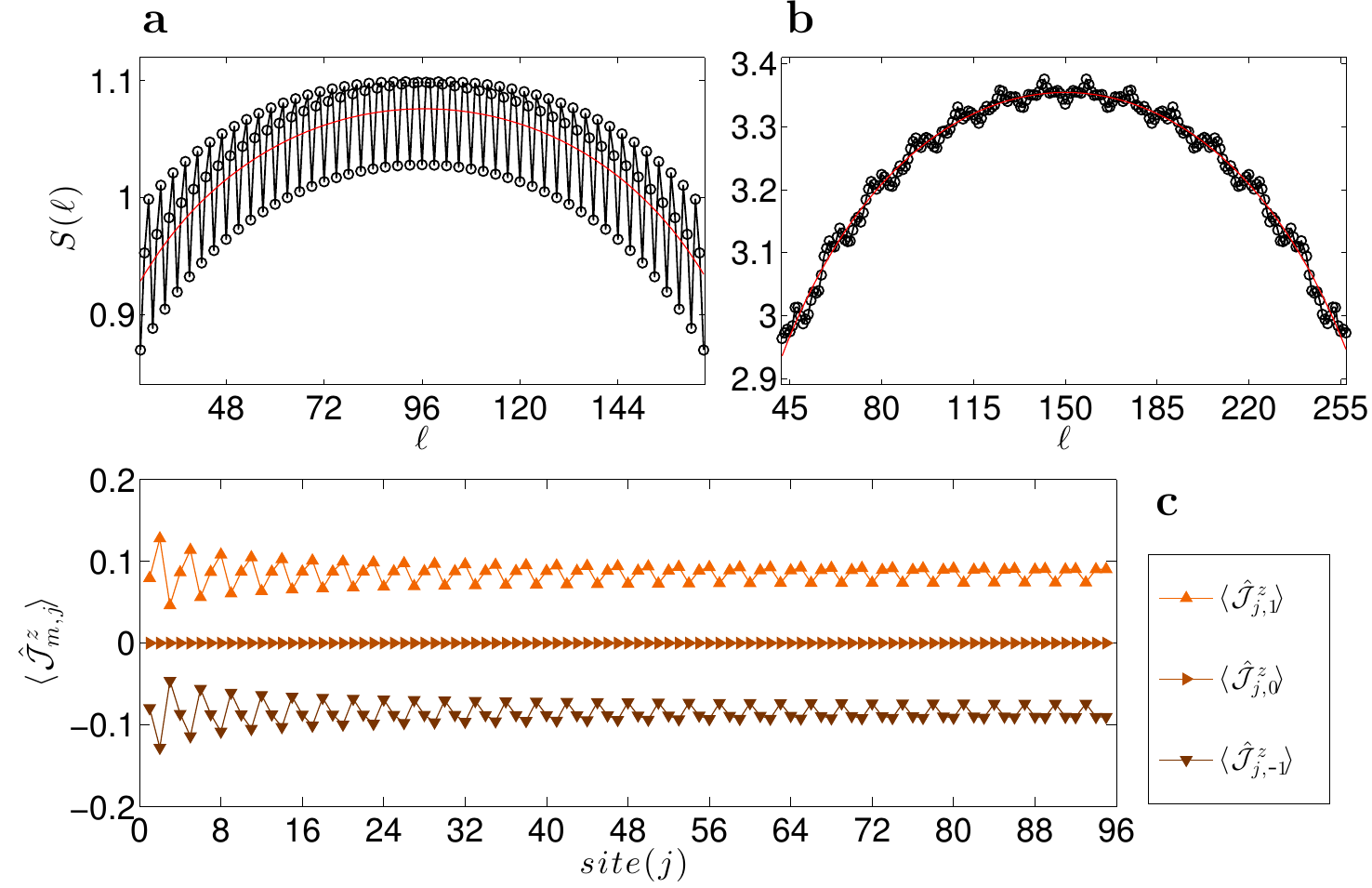}

 \caption{\textit{Helical liquids}. DMRG simulations of the entanglement entropy of the state for $I=1$, $k_{\rm SO} = \pi/3$, $\nu = \frac 23$, $\Omega / t = 1$ for 
 $U/t \to \infty$, $L=192$ ({\textbf{a}}) and $U/t = 0$, $L=300$ ({\textbf{b}}). Thin red lines are fits with the Calabrese-Cardy formula which yield $c=1.0  \pm 0.2$ 
 ({\textbf{a}}) and $c = 2.94 \pm 0.06$ ({\textbf{b}}). ({\textbf{c}}) Helical currents $\langle \hat {\mathcal J}^z_{j,m}\rangle$ for the case $\Omega = t$.
 For clarity, only the first half of the chain is plotted. 
 }
\label{fig:Helical:OBC}
\end{figure}

The simultaneous presence of two gapless modes ($c=1$) and of helical current patterns, indicating the presence of a helical 
state, results from the interplay between interactions and the gauge potential. In the non-interacting case, at this filling 
the system has three gapless  edge modes, which reflect in a central charge $c=3$, as shown in Figure~\ref{fig:Helical:OBC}. 
On the other hand, if we tune $\Omega=0$ (and thus $k_{\rm SO}$ disappears from the Hamiltonian), the model is equivalent to a SU($3$)
Hubbard model, which cannot develop helical currents, since the Hamiltonian is real. Even the milder request $\Omega \neq 0$,
$k_{\rm SO}=0$ results in a state without helical currents for the same reason.

Finally, we would like to comment on the role of $\hat {\mathcal H}_2$ in the $I\geq1$ case: 
the presence of the multi-photon transition leads to the absence  of gapless fractional helical modes and vice versa. 
Upon interpreting our setup as a spinless system with one synthetic dimension~\cite{Boada_2012, Celi_2014} and boundary conditions 
depending on $\hat {\mathcal H}_2$,
we have a phenomenology reminiscent of that of topological models with edge modes, which gap out once the boundary conditions are changed.
In the non-interacting limit,
this analogy becomes exact for  fillings with $q=1$ in the $I \to \infty$ limit, where the physics of the integer QHE is recovered. 
It is remarkable that this analogy holds also 
for interacting systems ($q>1$) and for small $I$. This rich phenomenology is lost for $I=1/2$ (or for spin-$1/2$ quantum liquids, 
such as electron degenerate gases): 
in a two-leg geometry, there is no difference between a cylinder and a stripe.

\paragraph{ \textit{ Conclusions.} }

Alkaline-earth(-like) gases provide access to the physics of  (quasi-)two-dimensional models through the mapping of the nuclear-spin states 
to a  (finite) synthetic dimension~\cite{Boada_2012}. In the system considered in this article,
the synthetic lattice is pierced by a magnetic gauge potential with flux per plaquette $\Phi / \Phi_0 =2 k_{\rm SO}$ ($\Phi_0$ is 
the quantum of flux)~\cite{Celi_2014}, whose properties are thus related to the original SOC potential.

Interestingly, for $I \geq 1$, the gapped phases that we have studied 
are amenable to an intriguing interpretation by introducing an extra synthetic-dimension.
Indeed, when $\hat {\mathcal H}_2\neq 0$ the effective geometry of the system is that of a narrow cylinder since the 
synthetic-dimension length, $2I+1$, is much smaller than the real-space one.
In this limit, usually called the \textit{thin-torus limit} (TTL)~\cite{Bergholtz2008, Bernevig_2012}, the fractional quantum Hall states 
are (continuously) turned into charge-density waves, with features which coincide with those presented in this article.
First, the filling $\nu$ coincides with the ratio between the number of particles and the number of magnetic fluxes piercing 
the synthetic lattice, $N/N_{\Phi}$, 
which is the well-known condition for observing the Laughlin series in QHE. 
Second, the density waves found in the TTL for $\nu = 1/q$
display one particle every  $q$ sites, which is what we show in Fig.~\ref{fig:Charge:Magnetic}.
Additionally, the helical properties which appear for $\hat {\mathcal H}_2 = 0$ can be interpreted as precursors of the  edge-modes 
of the QHE~\cite{Celi_2014}.

It is remarkable to observe the presence of these features in our system, since the  SU($2I+1$)-invariant interaction is, in the quasi-two-dimensional picture, strongly anisotropic, 
short-ranged along the chain and infinite-ranged in the  synthetic direction, and does not resemble the features neither of the Coulomb 
nor of the contact repulsion which are usually considered in the QHE theory. 

Spin-orbit coupled alkaline-earth(-like) gases are thus a natural quantum simulator of the physics of the fractional QHE in an array 
of quantum  wires~\cite{Kane_2002, Teo_2014}.
More precisely, because of the unusual properties of the interaction, we are dealing here with an unconventional form of QHE, and 
it is an exciting perspective to investigate up to which point it shares features with the standard QHE.
More generally, large-$I$ Fermi gases provide a valuable 
experimental toolbox for the study of two-dimensional exotic phases of matter
through coupled arrays of one-dimensional systems~\cite{Kane_2002, Teo_2014, Klinovaja_2014, Neupert_2014, Sagi_2014, Grusdt_2014}.
Additionally, the Raman coupling which we have described may realize a $\mathbb Z_{3}$-invariant model: 
a symmetry-class which supports topological phases with edge  parafermions~\cite{Fendley_2012, Milsted_2014}.

Concluding, in this article we have provided an analytical and numerical study of spin-orbit coupled alkaline-earth(-like) gases and demonstrated
the existence of a full hierarchy of gapless and gapped phases with exotic properties.
In particular, we have identified and  characterised the phases which can be experimentally realised with
state-of-the-art cold-atom experiments,
ranging from magnetic crystals to helical phases~\cite{Oreg_2014},
identifying the effect of the simultaneous presence of interactions and gauge potentials.
Our study also suggests that
alkaline-earth(-like) atoms are  a promising tool for bringing ultra-cold atomic gases into the quantum Hall regime, a 
long-standing and yet to be achieved goal, through the access of its thin-torus limit.
The unprecedented versatility of these setups motivates further speculations and research: we leave for example as an interesting open 
perspective the extension of this study to multicomponent bosonic systems~\cite{Dhar_2013, Piraud_2014, Wel_2014, Tokuno_2014, Hugel_2014, Petrescu_2015, Keles_2015}.

\paragraph{ \textit{Acknowledgements.}}

We gratefully acknowledge enlightening discussions with 
A. Celi, M. Burrello, M. Dalmonte, 
F. Gerbier, S. Manmana, M. Rizzi, E. Sela, A. Stern, H.-H. Tu and D. Vodola.
We particularly thank the experimental group at LENS in Florence: G. Cappellini, J. Catani, L. Fallani, L. Livi,  S. Mancini, G. Pagano and C. Sias for the discussion of the experimental details.
L.T. acknowledges financial support from the EU integrated project SIQS. D.R. and L.M. acknowledge the Italian MIUR through FIRB project RBFR12NLNA.
L.M. acknowledges Regione Toscana POR FSE 2007-2013.



\clearpage

\newpage

\clearpage 
\setcounter{equation}{0}%
\setcounter{figure}{0}%
\setcounter{table}{0}%
\renewcommand{\thetable}{S\arabic{table}}
\renewcommand{\theequation}{S\arabic{equation}}
\renewcommand{\thefigure}{S\arabic{figure}}

\setcounter{page}{1}

\begin{center}
{\Large Supplementary Material

\vspace{0.3cm}

\MyTitle}
\end{center}

\paragraph{ \textit{Relation with Rashba SOC.} } 

Let us first explicitly display that the unitary transformation $\hat {\mathcal U}$ defined in the Results leads to a Hamiltonian which is formally equivalent to a Rashba SOC model:
\begin{align}
\hat {\mathcal U} \, [
\hat {\mathcal H}_0+\hat {\mathcal H}_1 ] \, 
\hat {\mathcal U}^\dagger 
=- t \sum_{j,m} 
\left( e^{i 2 k_{\rm SO} m} \hat c_{j, m}^\dagger \hat c_{j+1, m} + \mathrm {H.c.} \right)+ \nonumber \\
+ \hat {\mathcal H}_{\rm int} 
+\Omega \sum_{j} \hspace{-0.2cm} \sum_{m = -I}^{I-1} \left( g_m \hat c_{j, m}^\dagger  \hat c_{j, m+1} +\mathrm {H.c.} 
\right). \nonumber
\end{align}
We have already commented on the fact that the term proportional to $t$ is the lattice version of a Rashba SOC. Concerning interactions,
it is natural to assume that $\hat {\mathcal H}_{\rm int}$ depends only on the density operators $\hat {n}_{j,m}$, which are left unchanged by $\hat {\mathcal U}$; thus: 
$\hat {\mathcal U} \,
\hat {\mathcal H}_{\rm int} \,
\hat {\mathcal U}^\dagger =\hat {\mathcal H}_{\rm int} $.
Finally, the term proportional to $\Omega$ is the result of a magnetic field applied along the $\hat x$ direction, perpendicular thus to the quantization axis of the SOC.
In an open chain, such as in our numerical simulations, these avoids the possibility of gauging away the phase $2 k_{\rm SO}$.

\paragraph{ \textit{Bosonization.}} 

In the following we briefly discuss the existence of the magnetic crystals within the bosonization framework.
The discussion follows the guidelines set in Ref.~\cite{Kane_2002, Oreg_2014, Teo_2014}.
To this aim we formally rewrite the fermionic operator as
\begin{eqnarray}
	\hat c_m(x)=\sum_{n=0} q_n \left(e^{i(2n+1)k_Fx} (\hat \psi^\dagger_{-,m} \hat \psi_{+,m})^n \hat \psi_{+,m}+\right. \nonumber\\
	\left.+e^{-i(2n+1)k_Fx}  (\hat \psi^\dagger_{+,m} \hat \psi_{-,m})^n \hat \psi_{-,m} \right) \label{haldane}
\end{eqnarray} 
in order to take into account the non-linearities of the free spectrum which play a non trivial role when interactions are strong enough; $q_n$ are unknown coefficients. 
Applying~\eqref{haldane} to a single term $e^{-2ik_{\rm SO}x}\; \hat{c}^\dagger_m(x) \hat{c}_{m+1}(x)$ of $\hat {\mathcal H}_1$ or to $\hat {\mathcal H}_2$ we get several contributions, of which those in the form:
\begin{equation}
 e^{2i[k_F(n+{n'}+1)-k_{\rm SO}]x} 
\; (\hat \chi^\dagger_m)^n 
\; \hat \psi^\dagger_{+,m} \hat \psi_{-,m+1} 
\; (\hat \chi^\dagger_{m+1})^{n'}; \label{yy1} 
\end{equation}
are particularly interesting.
Here, we have introduced the notation: $\hat \chi_m \equiv \hat \psi^\dagger_{-,m} \hat \psi_{+,m}$; $m=-I,...,I$ (with $I+1 \equiv -I$). 
The terms~\eqref{yy1} 
are the only ones which conserve momentum independently on the value of $k_{\rm SO}$, provided that
\begin{equation}
k_F=k_{\rm SO}/q, \quad \text{with} \quad q \equiv n+n'+1 \label{KF}.
\end{equation}
When $n=n'$ (and thus $q = 2n+1$ is an odd integer), 
the bosonized version of such operators is:
\begin{eqnarray}
\cos \left[n(\hat \phi_{+,m} + \hat  \phi_{-,m}+\hat \phi_{+,m+1}+\hat \phi_{-,m+1}) +\right. \nonumber\\
\left.\hat \phi_{+,m} + \hat \phi_{-,m+1} \right] \equiv \cos \hat{\mathcal{O}}_{m}. \label{jlo}
\end{eqnarray}
(see the Results section for the definition of the $\hat \phi_{\pm, m}(x)$).
Note that 
 $[\hat{\mathcal{O}}_{m}, \hat{\mathcal{O}}_{m'}]=0$ for $m,m'= -I, ..., I$.   
When the operators $\hat {\mathcal{O}}_{m}$  ($m=-I,...,I$) are relevant in the renormalization group sense 
(here we assume to be in the condition for which this is true),
they are pinned to minimize the quantity 
$\cos [\hat{\mathcal{O}}_{m}] $  and in a semi-classic approach, 
i.e. $\cos \hat {\mathcal{O}}_{m,m+1} \approx 1-\hat {\mathcal{O}}_{m,m+1}^2 $,  
the gapless Luttinger Hamiltonian $\hat {\mathcal H}_0$ acquires $2I+1$  mass gaps (i.e. it becomes fully gapped).

The existence of fractional phases with $q$ even cannot 
be straightforwardly explained using bosonization. 
Let us consider as an example the case $q=2$.
For each term  (\ref{yy1}) we can choose  $n=0$ and $n'=1$ 
but also $n=1$ and $n'=0$, so that 
$2(2I+1)$ sine-Gordon terms with non commuting arguments and with the same scaling dimension appear. In this case a semi-classic approach cannot be used and the solution of the resulting Hamiltonian is an interesting question.

As a last remark, we mention that
the bosonization framework can be used to show
 the existence of fractional insulating phases for $p>1$ and $q$ odd, 
 by introducing a fictitious coupling between different nuclear spin states of the form~\cite{Kane_2002}:
 \begin{eqnarray}
 \hat {\mathcal H}_1 + \hat {\mathcal H}_2 
= \Omega \sum_{j} \left[\sum_{m=-I}^{I-p} e^{-2i k_{SO} j} 
\hat c_{j, m}^\dagger  \hat c_{j, m+p} +\right. \nonumber\\
\left.+\sum_{z=1}^{p}
e^{-2i k_{SO} j} \hat c^{\dagger}_{j,I-p+z} \hat c_{j,-I+z+1}+\mathrm{H.c.}
 \right]
\end{eqnarray}
which models higher-order couplings between spin states with $\Delta m \neq \pm 1$. Bosonization of this term yields results which are completely equivalent to the previous ones, apart from the fillings at which gapped phases appear.

When we consider atoms
with two nuclear spin states, namely $I=1/2$, the Hamiltonian $\hat {\mathcal H}_{2}=0$. Gapped phases at fillings $\nu=1/q$
with odd $q$ are stabilised by~\eqref{yy1} and by
\begin{equation}
 e^{2i[k_F(-n-{n'}-1)-k_{\rm SO}]x} 
\;  (\hat \chi_m)^n 
\; \hat \psi^\dagger_{-,m} \hat \psi_{+,m+1} 
\; (\hat \chi_{m+1})^{n'}, \label{yyy1} 
\end{equation}
which also appear in the previous expansion.
With $m=-1/2$, they imply $k_{SO}=\pi/2$ and
$k_F=\pi/(2q)$ respectively. As already discussed in the main text the term (\ref{yyy1}) disappears in a continuum limit.  
If we introduce the  charge ($c$) and spin ($s$) bosonic fields $ \hat{\phi}_{r,\sigma}=  \; 1/\sqrt{2} [\hat{\varphi}_c - r\hat{\theta}_c + \sigma (\hat{\varphi}_s-r\hat{\theta}_s)]$
the Hamiltoninan $\hat {\mathcal H}_{0}$ can be written as
\begin{equation}
	\hat {\mathcal H}_{0}=  \sum_{\lambda=c,s} \frac{u_\lambda}{2\pi} \int dx \left(K_\lambda (\partial_x \hat{\theta}_\lambda)^2 
	+\frac{1}{K_\lambda} (\partial_x \hat{\varphi}_\lambda)^2 \right),
	\label{hamfree}
\end{equation}
$K_\lambda$ is the usual Luttinger parameter which takes into account the strength of the interaction,
 $u_\lambda=v_F/K_\lambda$ is the renormalized Fermi velocity. 
 On the other hand the terms (\ref{yy1}) and (\ref{yyy1})
\begin{equation}
\hat{\mathcal{H}}_{1} \propto \sum_{\eta=\pm}\int dx \cos[\sqrt{2}(q \hat{\varphi}_c +\eta \hat{\theta}_s)]\label{SG}
\end{equation}
couple the charge and the spin degrees of freedom. A renormalization-group calculation shows that they
are relevant when $K_c < 3/q^2$, assuming a SU(2) invariant interaction, i.e. $K_s=1$~\cite{Oreg_2014}. 
Physically, this means that an on-site repulsive 
interaction, for which $1/2 \leq K_c < 1$~\cite{Schulz_1990},
cannot stabilize a gapped phase for $q\geq 3$.
Longer-range interactions are thus necessary 
(for nearest neighbor repulsion, for instance, 
it is possible to achieve $K_c < 1/2$~\cite{Schulz_1990}), and in general
we expect that, even for $I \geq 1$, the higher
the value of $q$, the longer the range of the interactions
required to open the gap.

\paragraph{ \textit{Numerical simulations.}} 

DMRG is an algorithm which performs a search of the ground-state of a Hamiltonian in the space of matrix-product states (MPS), a class of states with finite correlations characterised by the so-called bond link, $D$~\cite{Schollwoeck_2005, Schollwoeck_2011}. 
In the limit $D=1$ MPS are product states, whereas for larger values of $D$ more quantum correlations can be described.

We consider chains with open boundaries and length comprised between $L =96$ and $L=192$; setting $D = 150$ we are able to describe the correlations in the states with sufficient accuracy. With these parameters
the effect of boundaries is irrelevant and
the errors on the observables are negligible on the scale of the symbols used in the figures. 
In the simulations of the gapped crystals convergence is helped by  the quantum numbers related to the conservation of each magnetization $\sum_j n_{j, \lambda}$; moreover we find that it is important to alternate the 
infinite-size version of the algorithm with the 
finite-size one.

\subsection*{Supplementary Material for the case $I=1/2$}

\begin{figure}
 
  \subfigure{\includegraphics[width=0.49\columnwidth,trim=0pt 0pt 0pt 0pt, clip]{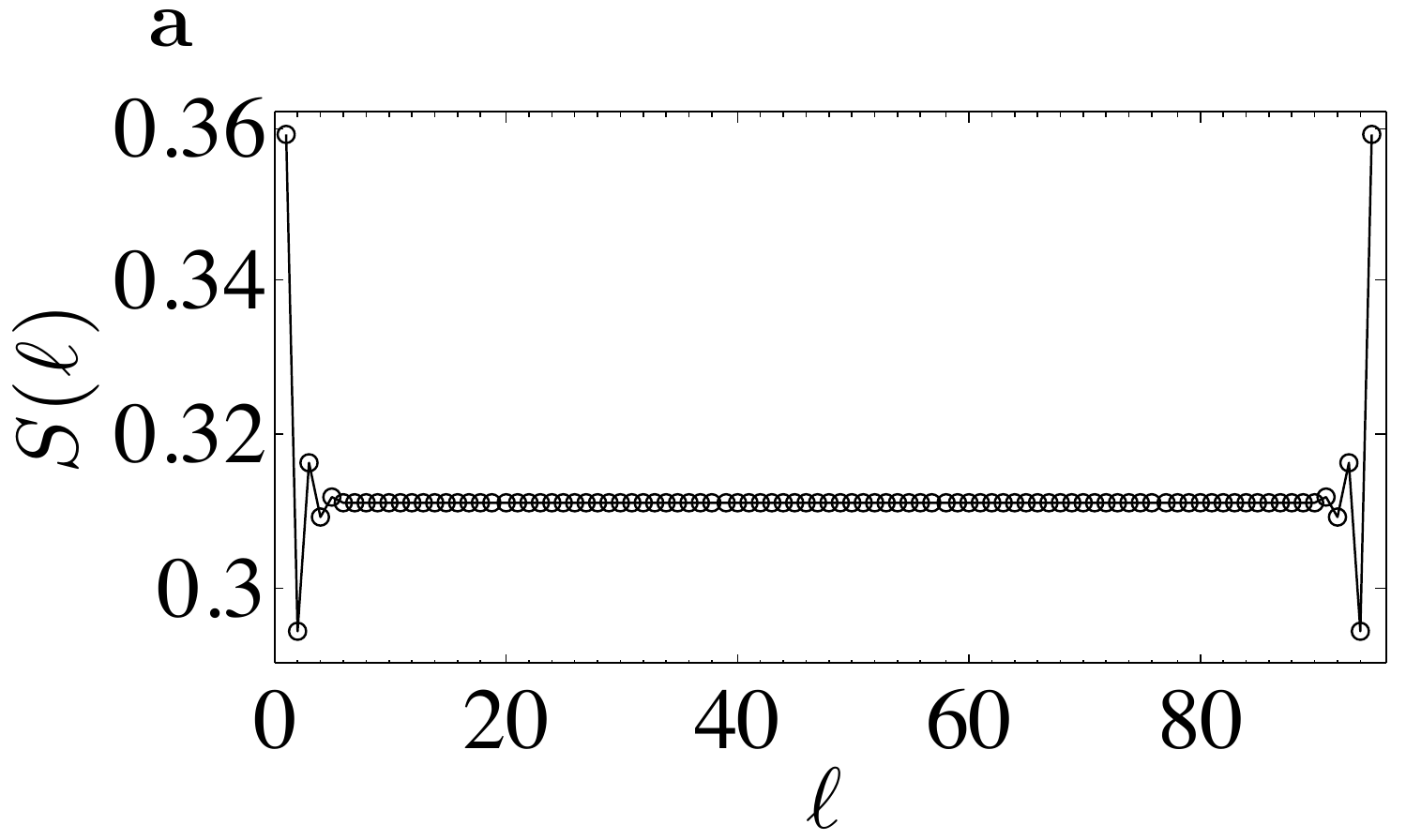}}
  \subfigure{\includegraphics[width=0.49\columnwidth,trim=0pt 0pt 0pt 0pt, clip]{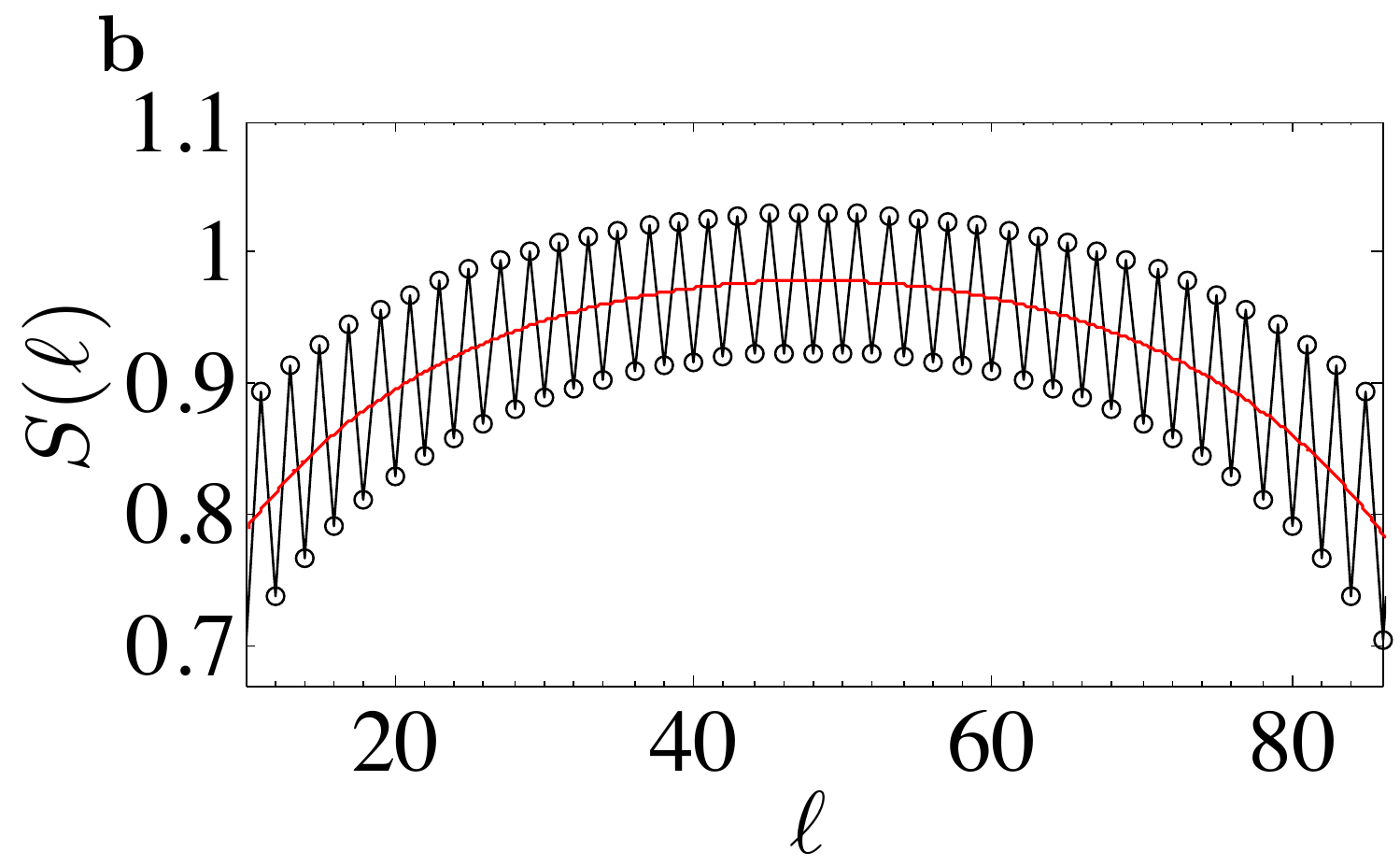}}
 
  \caption{Entanglement entropy $S(\ell)$ of the ground state of a SU($2$) fermionic gas for $k_{\rm SO}=\pi/2$ (\textbf{a}) and $k_{SO}=\pi/3$  (\textbf{b}); simulation parameters are $\nu=1/2$, $\Omega/t=1$ and $U/t\to\infty$. The fitted value of the central charge in panel  (\textbf{b}) is $c = 1.05 \pm 0.15$.
  }
  \label{EE:CORR}
 \end{figure}

\paragraph{ \textit{Numerical results for $k_{\rm SO}\neq \pi/2$.}} 
Let us numerically show that $k_{\rm SO}=\pi/2$ is a necessary condition to obtain crystalline phases for $I=1/2$. 
We consider as an example $\nu=1/2$ and the interaction
$\hat {\mathcal H} = U \sum_j \hat n_{j,m=1/2} \hat n_{j,m=-1/2}$
in the limit $U/t \to \infty$,
that stabilises a gapped phase for $k_{SO}=\pi/2$.
Consistently, the entanglement entropy of the ground state displays a clear area-law behaviour, see Fig.~\ref{EE:CORR}(\textbf{a}).
If $k_{\rm SO}$ is rather tuned to $\pi/3$, under the same conditions the entanglement entropy shows a non-area-law behaviour, which can be fitted with the Calabrese-Cardy formula, signaling the existence of a critical (gapless) ground state, see Fig.~\ref{EE:CORR}(\textbf{b}).

 \paragraph{ \textit{Gapping mechanism.}} 
 We now discuss the nature of the gapping mechanism responsible for the magnetic crystals studied in the text. 
 The combined action of the Raman coupling $\hat {\mathcal H}_{1}$ with $\hat {\mathcal H}_{\rm int}$  gives rise to two commuting Sine-Gordon terms:
 \begin{equation}
	 \int dx\ 	\cos[\sqrt{2}(q \hat{\phi}_c +\hat{\theta}_s)] +\int dx\ \cos[\sqrt{2}(q \hat{\phi}_c -\hat{\theta}_s)]. \label{mf}
	\end{equation} 
Additionally, it is known that when  $\nu = 1, \frac 12, \frac 13,...$ the bare interaction $\hat {\mathcal H}_{\rm int}$ leads to the additional Mott terms 
	\begin{equation}
	\hat{\mathcal{H}}_{\rm Mott,\nu} \propto \cos \left(\sqrt{8}\nu^{-1} \hat\phi_c\right),  \label{mott}
	\end{equation}
which constitute a gapping mechanism for the charge degrees of freedom  when  $ K_c <\nu^2$.
We now discuss the interplay of the terms~\eqref{mf} and~\eqref{mott} for different fillings $\nu$.

$\nu=1$.

\begin{figure}
  \includegraphics[width=0.85\columnwidth,trim=0pt 0pt 0pt 0pt, clip]{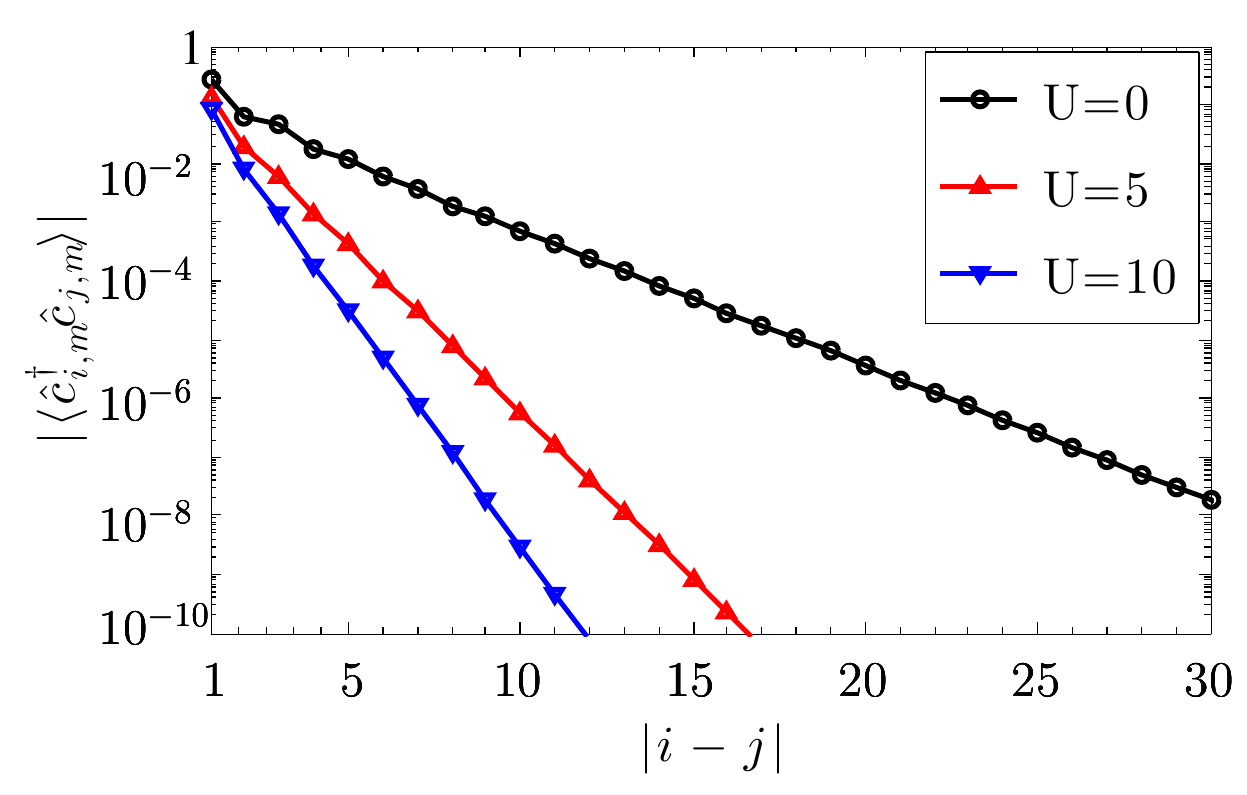}
  \caption{Single-particle correlator $|\langle \hat{c}^\dagger_{i,m} \hat{c}_{j,m}\rangle|$
  for different values of the interaction strength $U$; $\nu=1$, $\Omega/t=1$.}
  \label{fig:corr:SU2}
 \end{figure}

When $U=0$ and $\Omega \neq 0$ the ground state is fully gapped (band insulator)
due to the presence of the terms~\eqref{mf}  with $q=1$ (which are relevant for $K_c<3$). 
If $U>0$, the additional Mott term~\eqref{mott} with $\nu=1$ (which is relevant for $K_c<1$) has to be taken into account. 
As the arguments of these three terms commute, \eqref{mott} cannot modify the nature of the gapped phase or induce a phase transition
(note also that the terms~\eqref{mf} are more relevant). 
Furthermore, for $U>0$ we have numerically checked that the gap induced by the terms~\eqref{mf} is enhanced with respect to the non-interacting regime~\cite{Braunecker_2010}, as we can see by studying the correlation length $\xi$ associated to $ |\langle \hat{c}^\dagger_{i,m} \hat{c}_{j,m}\rangle| \sim e^{- |i-j|/\xi}$ (see Fig.~\ref{fig:corr:SU2})  for different values of $U$. 
The gapped phase is stabilised by $U$ since the correlation length $\xi$ decreases (and thus the gap is enhanced) when $U$ is increased.
The additional fact that
for $\Omega=0$ and $U\gg0$ the ground state is gapless (only charge degrees of freedom are gapped out)
asserts that we are not observing a standard Mott insulator.

$\nu=1/3$.
When $U=0$ and $\Omega \neq 0$ the ground state is gapless.  An incompressible phase is stabilised if $\Omega \neq 0$ and $V>0$ regardless the value of $U$. 
In this case we have: the terms~\eqref{mf} with $q=3$ (relevant if $K_c<1/3$) and the Mott term~\eqref{mott} with $\nu=1/3$ (relevant if $K_c<1/9$). Taking into account that $K_c>1/8$, since we only have a nearest-neighbour interaction~\cite{Schulz_1990}, 
we obtain that the Mott term is not relevant and it cannot constitute a gapping mechanism. 
We thus conclude that the incompressible phase we observe is due to the interplay of interactions and magnetic field only.
This is further supported by the observation that if we set $\Omega=0$ we can check numerically that the ground state is gapless and the entanglement entropy scales logarithmically with a central charge $c=2$ (implying four gapless modes) for all the possible values of $U$ and $V$.
 
 \subsection*{Supplementary Material for the case $I=1$}

\paragraph{\textit{Numerical results for $k_{\rm SO}=\pi/6$ and $\hat{\mathcal{H}}_2\neq0$.}} 

\begin{figure}[b]
	\includegraphics[width=\columnwidth]{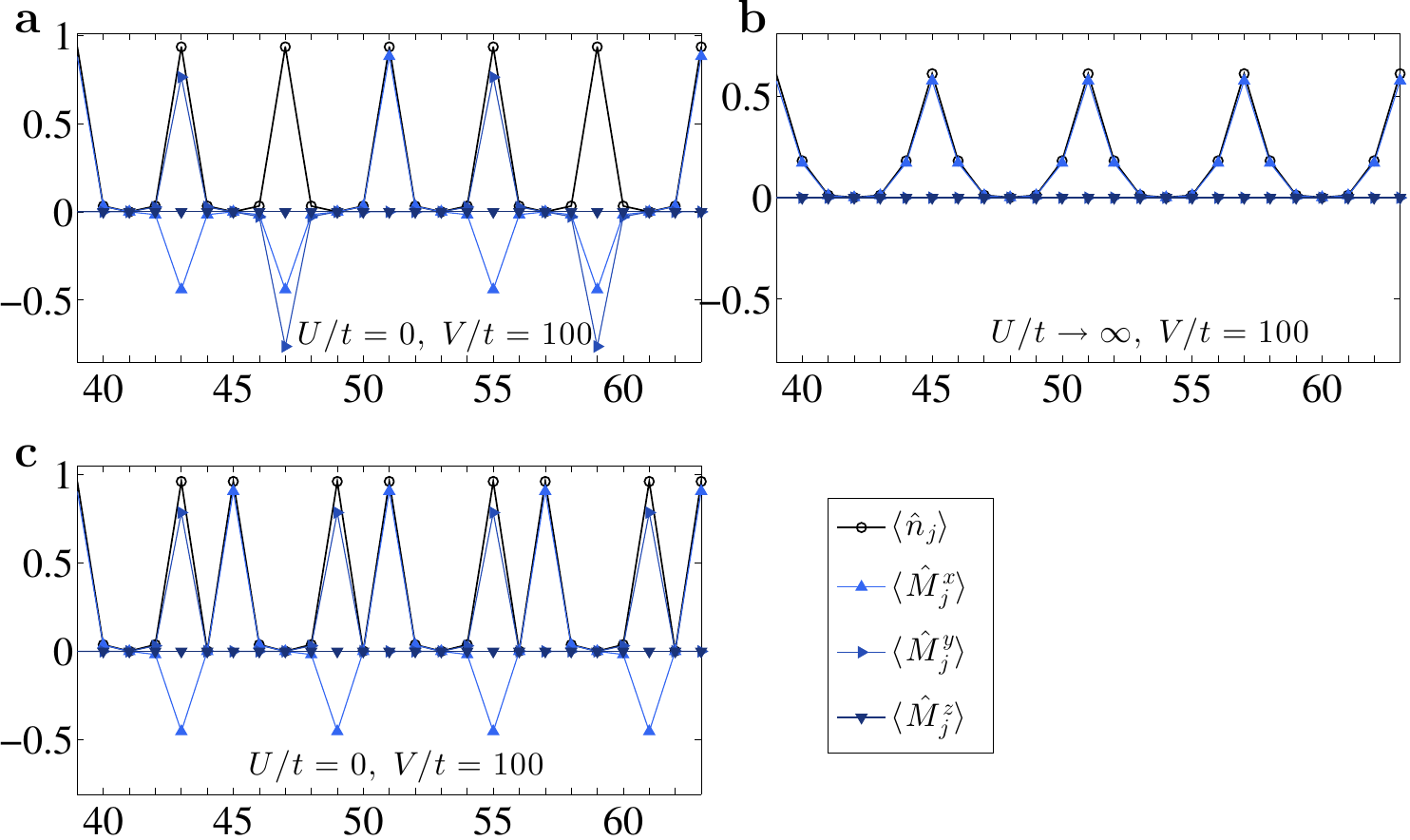}
	\caption{ Charge $\langle \hat n_i \rangle$ 
	and magnetic order $\langle \hat M^\alpha_i \rangle$ of the fractional phases at
	$\nu = \frac 12$ (\textbf{a}), $\frac 13$ (\textbf{b})
	and $\frac 23$ (\textbf{c}) for $k_{\rm SO} = \pi /6$
	as obtained from DMRG simulations 
	of a system of length $L=96$ with $\Omega/t = 1$.
	For the interaction parameters see the panels.
	Since the system is a crystal with small boundary effects, we only plot the central part of the system for a better readability.}
	\label{fig:Charge_Magnetic_SM}
\end{figure}

\begin{figure}
	\includegraphics[width=\columnwidth]{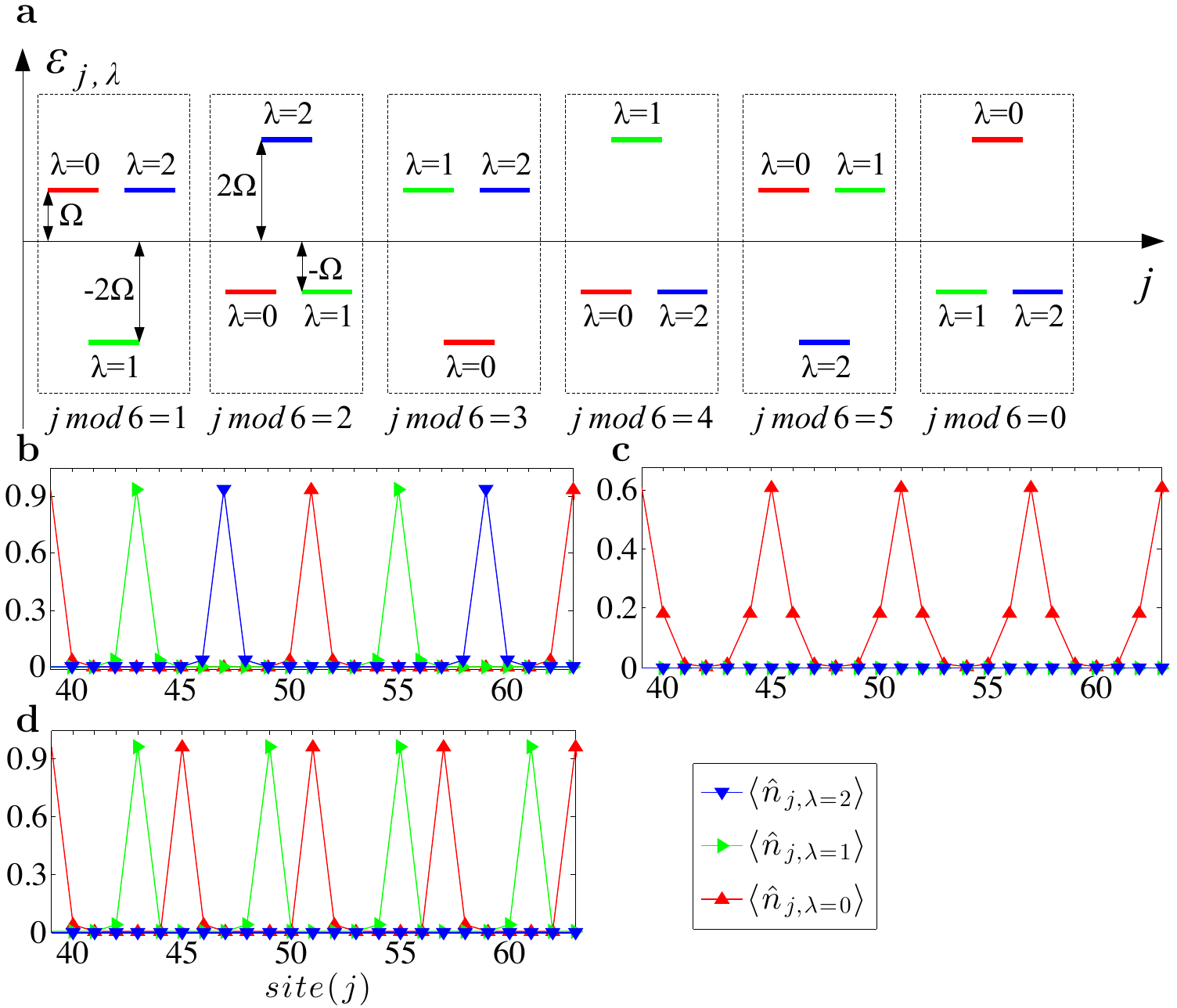}
	\caption{Energy-spin structure $\varepsilon_{j,\lambda}$ 
 	for $k_{\rm SO} = \pi/6$  (\textbf{a}).
  	Density plots $\langle \hat n_{j,\lambda} \rangle$ for fillings $\nu = \frac 12$ (\textbf{b}), 
 	$\frac 13$  (\textbf{c}) and $\frac 23$ (\textbf{d}) obtained through DMRG simulations (see the caption of Fig.~\ref{fig:Charge_Magnetic_SM} for the parameters).}
	\label{fig:SU3_PBC_SM}
\end{figure} 

In addition to the data for the fully-gapped phases for $k_{\rm SO}=\pi/3$ discussed in the main text, in Fig.~\ref{fig:Charge_Magnetic_SM} we present the density $\langle \hat n_j \rangle$ and magnetization profiles $\langle \hat M_j^\alpha \rangle$ of the gapped phases for $k_{\rm SO}=\pi/6$ and fillings $\nu=\frac12, \frac13, \frac23$. 
In Fig.~\ref{fig:SU3_PBC_SM}(\textbf{a}) we also show the landscape of the on-site eigenenergies $\varepsilon_{j,\lambda} = 2 \Omega \cos \left[ \frac{2 \pi \lambda} {2I+1} +2 k_{\rm SO} j \right]$ (note that they explicitly depend on $k_{\rm SO}$, and in the present case they display a six lattice site periodicity), in panels  (\textbf{b}-\textbf{d}) we show the density profiles in the rotated spin basis, $\langle \hat{n}_{j,\lambda} \rangle$.

\begin{figure}
 
  \subfigure{\includegraphics[width=0.85\columnwidth,trim=0pt 0pt 0pt 0pt, clip]{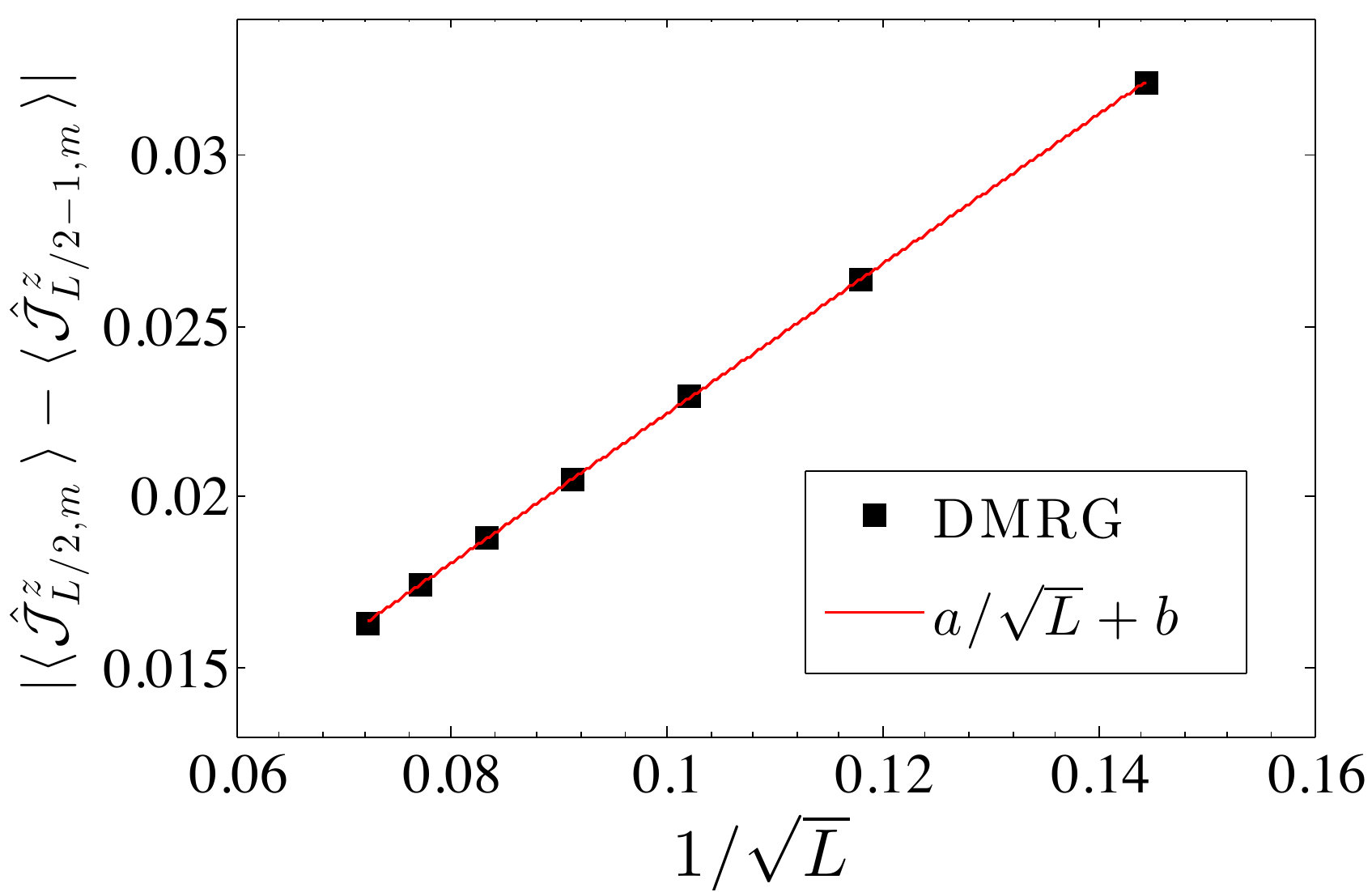}}
    \caption{The oscillations of the currents $\langle \hat {\mathcal J}_{j,m}^z  \rangle$ vanish in the thermodynamic limit; $a=(0.219\pm0.001)$ and $b=(5\cdot 10^{-5} \pm 4 \cdot 10^{-5})$.}
 \label{fig:SU3_osc_SM}
 \end{figure}

\begin{figure}
 
  \subfigure{\includegraphics[width=0.85\columnwidth,trim=0pt 0pt 0pt 0pt, clip]{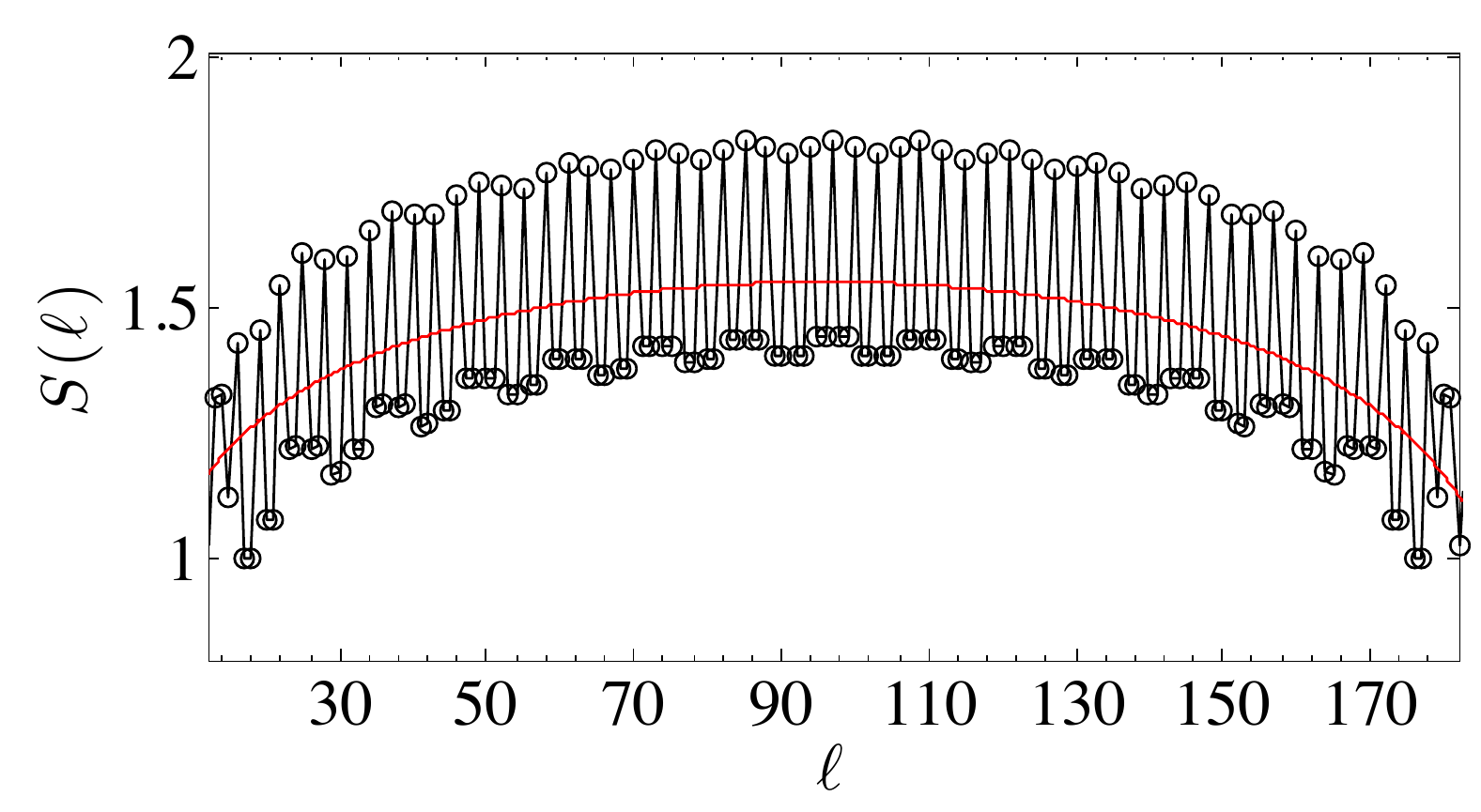}}
    \caption{DMRG simulation of the entanglement entropy of the state for $I=1$, $L=192$, $\nu = \frac 14$, $\Omega / t = 1$, $U/t \to \infty$ for
   $k_{\rm SO} = \pi/3$. Thin red line is a fit with the Calabrese-Cardy formula which yields $c=1.0  \pm 0.1$.
    }
 \label{fig:SU3_SM}
 \end{figure}

Remarkably, they resemble a \textit{diluted} version of the density profiles presented in the main text, suggesting that, at a fixed $\nu$, almost the same physics is obtained by scaling $k_{\rm SO}$ and the number of fermions by the same factor. However, different (e.g. longer range) interactions may be necessary to stabilize phases with the same $\nu$.

\paragraph{\textit{Numerical results for $k_{\rm SO}=\pi/3$ and $\hat{\mathcal{H}}_2=0$.}}  We show that the oscillations of the chiral currents vanish in the thermodynamic limit, and can therefore be interpreted as a boundary/finite-size effect (see Fig.~\ref{fig:SU3_osc_SM}). Conversely, the bulk value of such currents is independent from the system size.

\paragraph{\textit{Gapless phases with $\hat{\mathcal{H}}_2\neq0$.}} Finally, we show that the range of the interactions is an essential ingredient to stabilise fully-gapped phases when $\hat{\mathcal{H}}_2\neq0$. Fig.~\ref{fig:SU3_SM} demonstrates that a simple contact interaction cannot stabilise a gapped phase at $\nu=1/4$ with $k_{\rm SO}=\pi/3$: indeed, the ground-state von Neumann entanglement entropy displays a dependence on the subsystem size $\ell$ which is typical of a gapless phase.

\end{document}